\documentclass[11pt]{article}
\usepackage{mathrsfs}
\usepackage{amsmath}
\usepackage{amssymb}
\usepackage{cases}
\usepackage{ulem}
\usepackage{color}
\usepackage{tabularx}
\usepackage{cite}

\newcommand{\C}{ {\mathcal C}}
\newcommand{\R}{ {\mathcal R}}
\newcommand{\M}{ {\mathcal M}}
\newcommand{\Z}{ {\mathcal Z}}

\newcommand{\bF}{ {\mathbb F}}

\newcommand{\EOP} { \hfill $\Box$ }

\newcommand{\pf} { {\rm {\bf Proof.}} }

\newtheorem{theorem}{Theorem}[section]
\newtheorem{prop}[theorem]{Proposition}
\newtheorem{corollary}[theorem]{Corollary}
\newtheorem{lem}[theorem]{Lemma}

\newtheorem{remark}[theorem]{Remark}
\newtheorem{define}[theorem]{Definition}
\newtheorem{example}[theorem]{Example}

\begin{document}
\title{ New Constructions of Optimal Cyclic $(r, \delta)$ Locally Repairable Codes from Their Zeros }

\author{ \scriptsize Jing Qiu$^{1}$ \quad Dabin Zheng$^{1}$\footnote{Corresponding author \quad E-mail: dzheng@hubu.edu.cn } \quad
\quad  Fang-Wei Fu$^{2}$ \\
\scriptsize 1 \,\, Hubei Key Laboratory of Applied Mathematics, \\
\scriptsize Faculty of Mathematics and Statistics, Hubei University, Wuhan 430062, China \\
\scriptsize 2\,\, Chen Institute of Mathematics and LPMC, Nankai University, Tianjin 300071, China  \\
}
\date{}
\maketitle

\begin{abstract}
An $(r, \delta)$-locally repairable code ($(r, \delta)$-LRC for short) was introduced by Prakash et al.~\cite{Prakash2012} for tolerating multiple failed nodes in distributed storage systems, which was a generalization of the concept of $r$-LRCs produced by Gopalan et al.~\cite{Gopalan2012}. An $(r, \delta)$-LRC is said to be optimal if it achieves the Singleton-like bound. Recently, Chen et al.~\cite{Chen2018} generalized the construction of cyclic $r$-LRCs proposed by Tamo et al.~\cite{Tamo2015,Tamo2016} and constructed several classes of optimal $(r, \delta)$-LRCs of length $n$ for $n\, |\, (q-1)$ or $n\,|\, (q+1)$, respectively in terms of a union of the set of zeros controlling the minimum distance and the set of zeros ensuring the locality. Following the work of~\cite{Chen2018,Chen2019}, this paper first characterizes $(r, \delta)$-locality of a cyclic code via its zeros. Then we construct several classes of optimal cyclic $(r, \delta)$-LRCs of length~$n$ for $n\, |\, (q-1)$ or $n\,|\, (q+1)$, respectively from the product of two sets of zeros. Our constructions include all optimal cyclic $(r,\delta)$-LRCs proposed in~\cite{Chen2018,Chen2019}, and our method
seems more convenient to obtain optimal cyclic $(r, \delta)$-LRCs with flexible parameters. Moreover, many optimal cyclic $(r,\delta)$-LRCs of length $n$ for $n\, |\, (q-1)$ or $n\,|\, (q+1)$, respectively such that $(r+\delta-1)\nmid n$ can be obtained from our method.
\end{abstract}




\section{Introduction}

Let $\bF_q$ be a finite field with size $q$ and $\C$ an $[n, k]$ linear code over $\bF_q$. The $i$-th code symbol of  $\C$ is said to have $r$-{\it locality} $(1\leq r \leq k)$ if it can be recovered by accessing at most $r$ other code symbols in $\C$, i.e., the $i$-th code symbol can be expressed as a linear combination of $r$ other code symbols. If all the code symbols of $\C$ have locality $r$, then $\C$ is called an $r$ {\it locally repairable code} ($r$-LRC for short). This concept was introduced firstly by Gopalan et al.~\cite{Gopalan2012} for application of coding techniques to distributed storage systems. It was proved in~\cite{Gopalan2012,Tamo2014} that the minimum distance of an $r$-LRC~$\C$ is upper bounded by
\begin{equation}\label{eq:rbound}
d(\C) \leq n-k -\left\lceil \frac{k}{r} \right\rceil +2 .
\end{equation}
This bound is called the Singleton-like bound for LRCs. The linear codes meeting the above bound~(\ref{eq:rbound}) are called optimal $r$-LRCs.

In order to deal with the situation that multiple node failures occur in a distributed storage system, Prakash et al.~\cite{Prakash2012} introduced the concept of $(r, \delta)$-locality of linear codes, where $\delta\geq 2$, which generalized the notion of $r$-locality. The $i$-th code symbol of $\C$ is said to have $(r, \delta)$-locality ($\delta \geq 2$), if there exists a subset $S_i \subset \{1, 2, \ldots, n\}$ such that $i\in S_i$, $|S_i|\leq r+\delta-1$ and the punctured code $\C|_{S_i}$ has the minimum distance $d(\C|_{S_i}) \geq \delta$. The code $\C$ is said to have $(r, \delta)$-locality or be an $(r,\delta)$-LRC if all the code symbols have $(r, \delta)$-localities. A Singleton-like bound for the minimum distance of an $(r, \delta)$-LRC is given as follows~\cite{Prakash2012}:
\begin{equation}\label{eq:rdeltabound}
d(\C) \leq  n-k- \left(\left\lceil \frac{k}{r}\right\rceil -1 \right)(\delta-1) +1.
\end{equation}
Linear codes meeting this bound~(\ref{eq:rdeltabound}) are called optimal $(r, \delta)$-LRCs. Note that the notion of $r$-locality is a special case of the notion of $(r,\delta)$-locality for $\delta=2$.
In this case, the minimum distance bound~(\ref{eq:rdeltabound}) becomes to the bound~(\ref{eq:rbound}).

In the last decade, many constructions of optimal LRCs have been proposed, for example see~\cite{Chen2018,Chen2019,Fang2018,Gopalan2012,Goparaju2014,Hao2016,Jin2017,Jin2019,Kim2018,Li2019,Luo2019,Prakash2012,Song2014,Silberstein2013,
Sun2019,Tamo2014,Tamo2015,Tamo2016,Tan2019,Zeh2015} and references therein. A breakthrough construction of optimal LRCs was proposed by Tamo and Barg in~\cite{Tamo2014} via a generalization of the classical construction of Reed-Solomon codes. Along this way, Tamo et al.~\cite{Tamo2015,Tamo2016} constructed a family of optimal cyclic $r$-LRCs of lengths $q-1$ and its factors in terms of a union of the set of zeros controlling the minimum distance and the set of zeros ensuring the locality. Then their method was generalized to the $(r, \delta)$-locality case by Chen et\ al.\cite{Chen2018} and they also constructed many optimal cyclic $(r, \delta)$-LRCs with lengths $q+1$ and its factors by application of Berlekamp-Justesen codes. Very recently, many authors focused on constructions of optimal LRCs with unbounded length, for example see~\cite{Luo2019,Fang2018,Li2019,Sun2019,Jin2019}. Among all known constructions of optimal cyclic $(r, \delta)$-LRCs, the condition of $(r+\delta-1)\mid n$ has to be hold. To the best of our knowledge, there is no result about the existence or construction of optimal cyclic $(r,\delta)$-LRCs such that $(r+\delta-1)\nmid n$.

Inspired by the work of~\cite{Chen2018,Chen2019}, this paper aims to construct optimal cyclic $(r, \delta)$-LRCs in terms of
the product of two sets of zeros of cyclic codes. First, we characterize $(r, \delta)$-locality of cyclic codes from their zeros and show that
a cyclic code has $(r, \delta)$-locality if its complete defining set contains a product of two sets of zeros (see Theorem~\ref{thm:locality}).
By application of this result we propose several constructions of optimal cyclic $(r, \delta)$-LRCs of length $n$ for $n\,|\,(q-1)$ or $n\, |\, (q+1)$, respectively from the product of two sets of zeros of cyclic codes. All optimal cyclic $(r, \delta)$-LRCs proposed in~\cite{Chen2018,Chen2019} can be reconstructed from our method. It seems more convenient to obtain optimal cyclic $(r, \delta)$-LRCs with flexible parameters. Moreover, many optimal cyclic $(r,\delta)$-LRCs of length $n$ for $n\,|\,(q-1)$ or $n\, |\, (q+1)$, respectively such that $(r+\delta-1)\nmid n$ can be obtained from our constructions.

The rest of the paper is organized as follows. In Section~\ref{sec:preli}, we review some preliminaries on cyclic codes and some known constructions of optimal cyclic $(r,\delta)$-LRCs. Section~\ref{sec:locality} characterizes $(r, \delta)$-locality of cyclic codes from their zeros. In Section~\ref{sec:q-1}, we construct optimal cyclic $(r, \delta)$-LRCs of lengths $q-1$ and its factors in terms of the product of two sets of zeros. Section~\ref{sec:q+1} constructs optimal cyclic $(r, \delta)$-LRCs of lengths $q+1$ and its factors in terms of the product of two sets of zeros. Finally, Section~\ref{sec:concluding} concludes this paper.

\section{Preliminaries }\label{sec:preli}

\subsection{Cyclic codes and their complete defining sets}\label{subsec:def}

An $[n, k]$ cyclic code $\C$ over the finite field $\bF_q$ is a $k$-dimensional linear subspace of $\bF_q^n$ satisfying the condition that $(c_{n-1}, c_0, \cdots, c_{n-2})\in \C$ whenever $(c_0, c_1, \cdots, c_{n-1})\in \C$. It is well known that any cyclic code $\C$ of length $n$ over $\bF_q$ corresponds to an ideal of $\bF_q[x]/(x^n-1)$, and can be expressed as $\C=\langle g(x)\rangle$, where $g(x)$ is a monic polynomial over $\bF_q$ and $g(x)\,|\,(x^n-1)$. The $g(x)$ is called the generator polynomial and $h(x)=(x^n-1)/g(x)$ is referred to as the parity-check polynomial of $\C$~\cite{Macwilliams1977}. It is clear that the dimension of $\C$ equals $n-\deg(g(x))$.
 For any codeword $c=(c_0, c_1, ..., c_{n-1}) \in \C$, we correspond it to a polynomial $c(x)={\sum}_{i=0}^{n-1} c_i x^i \in \langle g(x)\rangle$. If $n$ and $q$ are co-prime, then $g(x)$ is uniquely determined by the set of its roots. The zeros of $g(x)$ are also called the zeros of $\C$, and $\Z_{\C}=\{\alpha^{\ell_{j}}|g(\alpha^{\ell_{j}})=0, j=1,2,\dots,n-k \}$ is called the \textit{complete defining set} of $\C$, where $\alpha$ is the $n$-th primitive root of unity in some extension of $\bF_q$.

In the next sections, adopt the following notation unless otherwise stated:
\noindent\begin{itemize}
\item $q$ is a power of a prime and $\bF_q$ is a finite field of size $q$.
\item $n$ is a positive integer with $(n,q)=1$, $d$ is the order of $q$ modulo $n$ and $\bF_{q^d}$ is the extension of $\bF_q$ with degree $d$.
\item $\alpha$ is a primitive $n$-th root of unity in $\bF_{q^d}$ and $\R_n$ is the set of all $n$-th roots of unity.
\item $AB$ is the set $\left\{ \beta\gamma \,\, |\,\, \beta \in A, \gamma \in B\right\}$, where $A, B$ are subsets of $\R_n$.
\end{itemize}

The operation of multiplying by $q$ divides the integers modulo $n$ into sets called $q$-cyclotomic cosets modulo $n$. For an integer $s$ with $0\leq s\leq n-1$,
the $q$-cyclotomic coset of $s$ modulo $n$ is defined by
\[ \left\{  s,\,\, sq,\,\, sq^2, \,\,\cdots, \,\,sq^{\ell_s-1} \right\}, \]
where $\ell_s$ is the smallest positive integer such that $sq^{\ell_s} \equiv s \,\, \,({\rm mod} \,\, n)$. Let $\C_A$ be a cyclic code over $\bF_q$ of length $n$
with the complete defining set $A$ for some $A\subset \R_n$. Then the set of exponents $j$ of $\alpha$ such that $\alpha^j \in A$, i.e.,
$\{ 0\leq j \leq n-1 \,\,|\,\, {\alpha}^j \in A \}$ is a union of some $q$-cyclotomic cosets modulo $n$, since the generator polynomial $\prod_{a \in A} (x-a)$
of $\C_A$ is a monic divisor polynomial over $\bF_q$ of $x^n-1$.

\begin{define}
A subset $\Omega=\{\alpha^{i_{1}},\alpha^{i_{2}},\dots,\alpha^{i_{\ell}}\}$ of $\R_{n}$ is called a consecutive set of length $\ell$
if a primitive $n$-th root $\beta$ of unity and an exponent $i$ exist such that $\Omega=\{ \beta^{i},\beta^{i+1},\dots ,\beta^{i+\ell-1}\}$.
\end{define}

The following lemmas are useful to establish our main results in this paper.

\begin{lem}\label{lem:BCH}\cite[BCH bound]{Macwilliams1977}
Let $\C$ be an $[n, k, d]$ cyclic code over $\bF_q$ and $\alpha$ a primitive $n$-th root of unity. If the complete defining set of $\C$
contains the following set
$$ \left\{ \alpha^u, \alpha^{u+b},\dots, \alpha^{u+(\delta-2)b} \right\},$$
where $b$ is a positive integer with $(b,n)=1$ and $u$ is a non-negative integer, then $d\geq \delta$.
\end{lem}

\begin{lem}\label{lem:BSbound}\cite[Betti-Sala bound]{Betti-Sala2006}
Let $u, m, \delta$ be non-negative integers with $m, \delta \geq 1$. Let $\C$ be an $[n, k, d]$ cyclic code over $\bF_q$ and $\alpha$
a primitive $n$-th root of unity. If the complete defining set of $\C$ contains the following set,
\[ \left\{  \alpha^u, \alpha^{u+1}, \dots, \alpha^{u+m\delta-1} \right\}  \bigcup_{i=0}^m \left\{ \alpha^{u+(m+i)\delta+1}, \alpha^{u+(m+i)\delta+2}, \dots, \alpha^{u+(m+i)\delta+\delta-1}\right\}, \]
then $d\geq m\delta+\delta$.
\end{lem}

As the generalization of the BCH bound, the minimum distance bound on cyclic codes proposed in~\cite{Betti-Sala2006} can be generalized
to the following case.
\begin{lem}\label{lem:genBSbound}
Let $u, m, b, \delta$ be non-negative integers with $m, b, \delta \geq 1$. Let $\C$ be an $[n, k, d]$ cyclic code over $\bF_q$ and $\alpha$ a primitive $n$-th root of unity. If the complete defining set of $\C$ contains the following set,
\[ \left\{  \alpha^u, \alpha^{u+b}, \dots, \alpha^{u+(m\delta-1)b}  \right\} \bigcup_{i=0}^m \left\{ \alpha^{u+((m+i)\delta+1)b}, \alpha^{u+((m+i)\delta+2)b}, \dots, \alpha^{u+((m+i)\delta+\delta-1)b}\right\}, \]
where $\gcd(b, n)=1$, then $d\geq m\delta+\delta$.
\end{lem}
\pf Let $\beta=\alpha^b$. Since $(b,n)=1$, $\beta$ is also a primitive $n$-th root of unity. So, $\alpha^{u}=\beta^{v}$ for some integer $v\in \{0,\dots,n-1\}$.
Then the complete defining set contains the following set:
\[ \left\{  \beta^v, \beta^{v+1}, \dots, \beta^{v+m\delta-1} \right\}\bigcup_{i=0}^m\left\{ \beta^{v+(m+i)\delta+1}, \beta^{v+(m+i)\delta+2}, \dots, \beta^{v+(m+i)\delta+\delta-1}\right\}.\]
The result follows from Lemma~\ref{lem:BSbound}.
\EOP

The above lemmas recall the results on the bound of the minimum distance of cyclic codes from their zeros. Next, we
further characterize the minimum distance of some cyclic codes in terms of their complete defining sets.

\begin{lem}\label{lem:distance}
Let $\C$ be an $[n, k]$ cyclic code over $\bF_q$ with the complete defining set $\Z\subset \R_n$, then the minimum distance of the dual code of $\C$ is equal to the minimum distance of the cyclic code with the complete defining set $\R_n\setminus \Z$.
\end{lem}
\pf  Let $g(x)$ denote the generator polynomial of $\C$ and $\Z$ its zero set. Then the set $\{ 0\leq  j\leq n-1 \,|\, \alpha^j \in \Z \}$ is a union of some $q$-cyclotomic cosets modulo~$n$. It is easy to see that the set $\{ 0\leq  j \leq n-1\, |\, \alpha^j \in \R_n \setminus \Z \}$ is also a union of some $q$-cyclotomic cosets modulo~$n$. Let $\C^\prime$ denote the cyclic code of length $n$ over $\bF_q$ with the complete defining set $\R_n\setminus \Z$. Then the generator polynomial of $\C^\prime$ is $h(x)=(x^{n}-1)/g(x)=\sum_{i=0}^{k}h_{i}x^{i}\in \bF_q[x]$. Let $\C^\perp$ denote the dual of $\C$ and it can be generated by
$h^{*}(x)=\sum_{i=0}^{k}h_{k-i}x^{i}\in \bF_q[x]$.

For any codeword $c^\prime(x)=\sum_{\ell=0}^{n-1}c^\prime_\ell x^\ell \in \C^\prime$ there exists $a(x)=\sum_{i=0}^{n-k-1} a_i x^i\in \bF_q[x]$  such that
$c^\prime(x) = a(x) h(x)$. So,
\begin{equation}\label{eq:cprime}
c^\prime_\ell = \sum_{i+j=\ell} a_i h_j,\,\, 0\leq i\leq n-k-1, \,\, 0\leq j \leq k .
\end{equation}
Let $a^{*}(x)=\sum_{i=0}^{n-k-1} a_{n-k-1-i} x^i$, then $c(x) = \sum_{\ell=0}^{n-1}c_\ell x^\ell =  a^{*}(x) h^{*}(x) \in \C^\perp$. So,
\begin{equation}\label{eq:c}
c_\ell = \sum_{i+j = n-1-\ell } a_i h_j, \,\, 0\leq i\leq n-k-1, \,\, 0\leq j \leq k .
\end{equation}
Comparing (\ref{eq:cprime}) and (\ref{eq:c}) we have $c^\prime_\ell = c_{n-1-\ell}$ for $\ell =0, 1, \ldots, n-1$.
In other words, for a codeword $(c_{0},c_{1}, \dots, c_{n-1})\in \C^\prime$, there always exists a codeword $(c_{n-1},c_{n-2},\dots,c_{0})\in C^{\perp}$,
and vice versa. This completes the proof.
\EOP

Below we give an example to illustrate Lemma 2.5.
\begin{example}
Let $q=2^3$, $n=7$ and $\alpha\in \bF_8$ be a primitive $7$-th root of unity. Let $\C$ and $\C^\prime$ be the cyclic codes over $\bF_8$ with the complete defining sets $\{ \alpha^3,\alpha^4,\alpha^5\}$ and $\{\alpha^0, \alpha^1, \alpha^2,\alpha^6\}$, respectively. Magma verifies that $d^\prime=d^\perp =4$, where $d^\prime$ and $d^\perp$ denote the minimum distance of $\C^\prime$ and the dual of $\C$, respectively. The experiment result is consistent with Lemma~\ref{lem:distance}.
\end{example}

\begin{prop}\label{prop:twod}
Let $\C_A$ denote the cyclic code of length $n$ over $\bF_q$ with the complete defining set $A\subset \R_n$. Let $\bar{\C}_A$ denote the cyclic
code of length~$n$ over $\bF_{q^d}$ with the same complete defining set $A$, i.e., $\bar{\C}_A$ is generated by the generator polynomial of $\C_A$ over $\bF_{q^d}$.
Then $d(\C_A)=d(\bar{\C}_A)$, where $d(\C_A)$ and $d(\bar{\C}_A)$ denote the minimum distance of $\C_A$ and $\bar{\C}_A$, respectively.
\end{prop}
\pf Since $\C_A$ is the subfiled subcode of $\bar{\C}_A$, $d(\C_A)\geq d(\bar{\C}_A)$. We only need to show $d(\C_A)\leq d(\bar{\C}_A)$.  Let $g(x) = \prod_{\beta\in A} (x-\beta)$ with degree $n-k$, which generate $\C_{A}$ over $\bF_q$ and $\bar{\C}_{A}$ over $\bF_{q^d}$, respectively, i.e.,
\[ \begin{split}
\C_{A}=\{g(x)f^\prime(x) | f^\prime(x)\in \bF_q[x], \deg f^\prime(x)\leq k-1 \},\\
\bar{\C}_A=\{g(x)f(x) | f(x)\in \bF_{q^d}[x], \deg f(x)\leq k-1 \}.
\end{split} \]
Since $\C_A$ is a cyclic code over $\bF_q$, $g(x)$ is a polynomial over $\bF_q$, and assume that $g(x)=\sum\limits_{i=0}^{n-k}g_{i}x^{i}\in \bF_q[x]$.

Suppose $c(x)=\sum\limits_{\ell=0}^{n-1}c_{\ell}x^{\ell}$ is a non-zero codeword of $\bar{\C}_A$ with the minimum Hamming weight~$d(\bar{\C}_A)$, and $c_{\ell} =0$ for $\ell \in I$, where $I\subset \left\{0,1,\dots, n-1\right\}$ and $|I| = n- d(\bar{\C}_A)$. There exists a non-zero polynomial $f(x)=\sum\limits_{j=0}^{k-1}f_jx^j \in \bF_{q^d}[x]$ such that
$c(x) = g(x) f(x)$. So,
\begin{equation}\label{eq:cl}
c_\ell = \sum_{i+j=\ell} g_i f_j,\,\, 0\leq i\leq n-k, \,\, 0\leq j \leq k-1, \,\, 0\leq \ell \leq n-1 .
\end{equation}
Let $\alpha_0,\alpha_1,\dots,\alpha_{d-1}$ be a basis of $\bF_{q^d}$ over $\bF_q$. Each coefficient $f_j$ of $f(x)$ can be represented as follows:
\begin{equation}\label{eq:fj}
f_j=\sum\limits_{t=0}^{d-1}f_{jt}\alpha_t, \,\,\, 0\leq j\leq k-1,
\end{equation}
where $f_{jt}\in \bF_q$ for $0\leq t\leq d-1$.
From assumption and equation (\ref{eq:cl}), for $\ell\in I$ we have
\begin{equation}\label{eq:cl1}
c_\ell= \sum\limits_{i+j=\ell} g_i f_j=0, \,\, \,\, 0\leq i\leq n-k, \,\, 0\leq j \leq k-1.
\end{equation}
Substituting $f_j$ in (\ref{eq:fj}) into (\ref{eq:cl1}) we get
\[\sum_{i+j=\ell} g_i f_j = \sum_{i+j=\ell} \left(\sum_{t=0}^{d-1}f_{jt}\alpha_t\right)g_i=\sum_{t=0}^{d-1}\left(\sum_{i+j=\ell}f_{jt}g_i\right)\alpha_t=0,\]
where $0\leq i\leq n-k, \,\, 0\leq j \leq k-1,\,\, \ell \in I$. So,
\begin{equation}\label{eq:cl2}
\sum\limits_{i+j=\ell} f_{jt} g_i=0, \,\,\,  0\leq  t\leq d-1, \,\,\, \ell \in I.
\end{equation}
Since $f(x)$ is non-zero, there is at least one non-zero $k$-tuple $(f_{0i}, f_{1i},\dots,f_{(k-1)i})$ for $0\leq i\leq d-1$. Assume that $(f_{0r}, f_{1r},\dots,f_{(k-1)r})\in \bF_q^k$ is a non-zero tuple and set $f^{'}(x)=\sum_{j=0}^{k-1}f_{jr}x^j\in \bF_q[x]$. From (\ref{eq:cl2}) we find a codeword $c^\prime(x)=\sum\limits_{\ell=0}^{n-1}c_{\ell}^\prime x^{\ell}=f^\prime(x)g(x)\in \C_A $ such that $c_\ell^\prime=0$ for $\ell \in I$. So, $d(\C_A)\leq d(\bar{\C}_A)$. This completes the proof.
\EOP

\begin{prop}\label{prop:exactd}
Let $n$ be a positive integer with $n\, |\, (q-1)$ and $\C$ be a cyclic code over $\bF_q$ with the complete defining set $\Z\subset \R_n$.
If $\Z$ contains a coset of a subgroup of $\R_n$ with order~$\ell$ and $\R_n\setminus \Z$ contains a consecutive set of length
$\frac{n}{\ell}-1$, then the minimum distance of the dual of $\C$ is exact $\frac{n}{\ell}$.
\end{prop}
\pf Let $\C^\perp$ and $d^\prime$ denote the dual of $\C$ and the minimum distance of $\C^\perp$, respectively. Let $s=\frac{n}{\ell}$ and $G=\langle \alpha^{s}\rangle$ be the subgroup of $\R_n$ of order $\ell$ and $\alpha^t G \subseteq \Z$ for some $t$ with $0\leq t\leq s-1$. Consider an $\ell\times n$ matrix as follows:
\[\M(\alpha^{t} G)=
\begin{pmatrix}
1 & \alpha^{t }       &  \alpha^{2t}     & \dots     & \alpha^{(n-1)t}     \\
1 & \alpha^{t+s}      &  \alpha^{2(t+s)} & \dots     & \alpha^{(n-1)(t+s)} \\
1 & \alpha^{t+2s}     &  \alpha^{2(t+2s)} & \dots    & \alpha^{(n-1)(t+2s)} \\
  \vdots & \vdots & \vdots & \vdots & \vdots  \\
1 & \alpha^{t+(\ell-1)s} &  \alpha^{2(t+(\ell-1)s)} &\dots & \alpha^{(n-1)(t+ (\ell-1)s)}
\end{pmatrix}.
\]
It is clear that each vector in the row space of $\M(\alpha^{t} G)$ is in~$\C^\perp$. Since
\[ \sum_{i=0}^{\ell-1} \alpha^{j (t + is)} = \alpha^{jt} \sum_{i=0}^{\ell-1} \alpha^{ijs} =\left\{
\begin{array}{lcl}
 \ell \alpha^{j t},  \quad {\rm if }\,\, \ell \,|\, j, \\ \\
 0, \quad {\rm otherwise },
 \end{array} \right.  \]
where $j=0, 1, \dots, n-1$. Adding up all rows in $\M(\alpha^t G)$ we get a codeword in $\C^{\perp}$ with Hamming weight~$s$.
On the other hand, there exists a consecutive set of length $s-1$ in $\R_n \setminus \Z$. By Lemma~\ref{lem:distance}
and Lemma~\ref{lem:BCH}, $d^\prime\geq s$. So, $d^\prime=s$.  \EOP

\subsection{Some known constructions of optimal cyclic $(r,\delta)$-LRCs }

Tamo et al.\cite{Tamo2015,Tamo2016} constructed a class of optimal cyclic $r$-LRCs  with lengths ${q-1}$ and its factors based on  Reed-Solomon codes. Immediately,
Chen et al.~\cite{Chen2018} generalized this construction to the case of cyclic $(r, \delta)$-LRCs. Moreover, they firstly constructed several classes of cyclic $(r, \delta)$-LRCs with lengths $q+1$ and its factors by application of Berlekamp-Justesen codes. In this section, we recall the construction of optimal cyclic $(r, \delta)$-LRCs with lengths ${q-1}$ and its factors in~\cite{Chen2018}, and the case of lengths $q+1$ and its factors is referred to~\cite{Chen2018,Chen2019}.

\begin{lem}\label{lem:bin2}\cite[Proposition~6]{Chen2018}
Let $n, r, \, \delta$ be positive integers such that $n\,|\,(q-1)$, $(r+\delta-1)|n$ and $\C$ be a cyclic code of length $n$ over $\bF_q$ with the complete defining set $\Z$. Let $0\leq l_1<l_2< \dots <l_{\delta-1}\leq r+\delta-2$ be an arithmetic progression with $\delta-1$ items and the common difference $b$, where $(b,n)=1$. If $\Z$ contains some cosets of the group of $\nu$-th roots of unity $\cup_{l}L_l$, where $L_l=\{\alpha^i|i\mod(r+\delta-1)=l \}$, $l=l_1,\dots, l_{\delta-1}$ and $\nu=n/(r+\delta-1)$, then $\C$ has $(r,\delta)$-locality.
\end{lem}

\begin{lem}\label{lem:construction-chen}~\cite[Construction~7]{Chen2018}
Let $r, \, \delta$ be positive integers such that $(r+\delta-1)|n$. Let $\alpha \in \bF_q$ be a primitive $n$-th root of unity, where $n\mid (q-1)$. Let $0 \leq l_1<l_2<\dots<l_{\delta-1}\leq r+\delta-2$ be an arithmetic progression with $\delta-1$ items and the common difference $b$, where $(b,n)=1$. Suppose $r\mid k$ and let
$\mu=\frac{k}{r}$. Consider the following sets of elements of $\bF_q$:
\[L_l =\{\alpha^i \,|\, i\mod (r +\delta - 1) = l\}, \ l= l_1, l_2, \dots, l_{\delta-1}\] and $D=\{\alpha^{t+eb} |e=0, 1,\dots,n-\mu(r+\delta-1)+\delta-2 \}$,
where $\alpha^t \in L_{l_1}$. Then the cyclic code $\C$ with the complete defining set $(\cup_l L_l ) \cup D$ is an optimal cyclic $(r,\delta)$-LRC with length $n$,
dimension $k$ and minimum distance $n-k+1-(\mu-1)(\delta-1)$.
\end{lem}

\section{$(r , \delta)$-locality of cyclic codes}\label{sec:locality}

It is known that any cyclic code $\C$ over $\bF_q$ has $r$-locality, where $r=d^\perp-1$, and $d^\perp$ is the minimum distance of the dual of $\C$.
In this section, we generalize this result and characterize the $(r,\delta)$-locality of cyclic codes in terms of their zeros. To this end, we first analyze the $(r,\delta)$-locality of cyclic codes from their parity check matrices.

\begin{lem}\label{lem:equdef}
Let $r, \delta$ be positive integers and $\C$ an $[n, k]$ cyclic code over $\bF_q$. Then $\C$ has $(r, \delta)$-locality if and only if there exists
a $t\times n$ matrix $H$ over $\bF_q$ with only $r+\delta-1\,\, (\leq n)$ non-zero columns whose rows are codewords in $\C^{\perp}$ such that any $\delta-1$
non-zero columns are linearly independent over $\bF_q$.
\end{lem}
\pf Assume that $H$ is a $t\times n$ matrix over $\bF_q$ with only $r+\delta-1$ non-zero columns whose rows are codewords in $\C^\perp$ such that any $\delta-1$ non-zero
columns of $H$ are linearly independent over $\bF_q$. Denote the index set of all non-zero columns of $H$ by $I$ and $|I|=r+\delta-1$. Let $\C|_I$ denote the punctured code
of $\C$ over the coordinate set $I$. If we can show that $d(\C|_I) \geq \delta$, then $\C$ has $(r,\delta)$-locality for the coordinates in $I$.
Since $\C$ is a cyclic code, each coordinate of the codeword has $(r,\delta)$-locality.
Let $\C(H)$ denote the linear code such that $H$ is the parity-check matrix. Let $H_I$ denote the submatrix consisting of columns of $H$ whose support is $I$.
Then $d(\C(H|_I)) \geq \delta$. It is easy to show that $\C|_I \subseteq \C(H_I)$. So, $d(\C|_I)\geq \delta$.

Conversely, let $I$ be a recover set for some coordinate $i\in [n]$ and $|I|=r+\delta-1$, where $[n] = \{ 1, 2, \dots, n\}$. Let $G$ be the generator matrix of $\C$,
then $G|_I$  generates $\C|_I$. Let $H^{\prime}$ be the matrix such that $H^\prime|_I$ is the parity-check matrix of $\C|_I$, and $H^\prime|_{[n]\setminus I}$ is a zero matrix. Then we have $H^\prime G^T=0$, so the rows of $H^\prime$ are codewords of $\C^\perp$, and the number of non-zero columns
of $H^\prime$ is $r+\delta-1$. Moreover, since $\C$ has $(r,\delta)$-locality, any $\delta-1$ non-zero columns of $H^\prime$ are linearly independent over $\bF_q$.
\EOP

In general, the zeros of a cyclic code of length~$n$ over $\bF_q$ are in the extension $\bF_{q^d}$ of $\bF_q$, where~$d$ is the order of $q$ modulo $n$. To characterize the
$(r, \delta)$-locality of cyclic codes in terms of their zeros, we still need the following lemma.

\begin{lem}\label{lem:localityofsubfiledsubcode}
 Let $\bar{\C}$ be an $[n, k]$ cyclic code over $\bF_{q^d}$ and $\C$ denote the subfield subcode of $\bar{\C}$ over $\bF_q$, where $d$ is the order of $q$ modulo $n$. If $\bar{\C}$ has $(r,\delta)$-locality, then $\C$ also has $(r,\delta)$-locality.
\end{lem}
\pf Since $\bar{\C}$ has $(r,\delta)$-locality, from Lemma~\ref{lem:equdef}, there exists a matrix $H=(a_{ij})_{t\times n}$ over $\bF_{q^d}$ with only $r+\delta-1$ non-zero columns whose rows are codewords in $\bar{\C}^{\perp}$ such that any $\delta-1$ non-zero columns are linearly independent over $\bF_{q^d}$, where $\bar{\C}^\perp$ denote the dual of $\bar{\C}$. Let $[a_{ij}]$ denote the column vector in $\bF_{q}^{d}$ corresponding to $a_{ij}$ for $i=1,2,\dots,t,\,\,j=1,2,\dots,n$. Then $[H] = \left( [a_{ij}]\right)_{td\times n}$ is a $td \times n$ matrix over $\bF_q$. For each codeword ${\bf c}\in \C$, we have $[H]\cdot {\bf c}^T=0$. So, each row of $[H]$ is a codeword of the dual of $\C$. Since $H$ has only $r+\delta-1$ non-zero columns and any $\delta-1$ columns among them are $\bF_{q^d}$-linear independent, $[H]$ also has only $r+\delta-1$ non-zero columns and any $\delta-1$ columns among them are $\bF_{q}$-linear independent. By Lemma~\ref{lem:equdef}, $\C$ has $(r,\delta)$-locality.
\EOP

\begin{theorem}\label{thm:locality}
Let $d$ be the order of $q$ modulo $n$, $A$ and $B$ be two subsets of $\R_n\subset \bF_{q^d}$ such that $d_A^\perp \geq d_B$, where $d_A^\perp$ and $d_B$ denote the minimum distance of the dual of the cyclic code of length~$n$ over $\bF_{q^d}$ with the complete defining set $A$ and the cyclic code of length~$n$ over $\bF_{q^d}$ with the complete defining set $B$, respectively. If the complete defining set of a cyclic code $\C$ of length~$n$ over $\bF_{q}$ contains $AB$, then $\C$ has $(d_{A}^{\perp}-d_B +1, d_B)$-locality.
\end{theorem}

\pf  Let $\bar{\C}$ be the cyclic code of length $n$ over $\bF_{q^d}$ generated by the generator polynomial of $\C$ over $\bF_{q}$.  It is easy to see that $\C$ is the subfield subcode of $\bar{\C}$. 
By Lemma~\ref{lem:localityofsubfiledsubcode} we only need to show that $\bar{\C}$ has $(d_{A}^{\perp}- d_B +1, d_B)$-locality.

Let $A=\{\alpha^{i_{1}}, \alpha^{i_{2}}, \dots, \alpha^{i_{u}}\}\subset \R_n$. Let $\bar{\C}_A^{\perp}$ and $d_{A}^{\perp}$ denote the dual of the cyclic code of length~$n$ over $\bF_{q^d}$ with the complete defining set $A$ and the minimum distance of $\bar{\C}_A^{\perp}$, respectively. Set
\[\M(A)=\begin{pmatrix}
1 & \alpha^{i_{1}} & \alpha^{2i_{1}} & \dots & \alpha^{(n-1)i_{1}}\\
1 & \alpha^{i_{2}} & \alpha^{2i_{2}} & \dots & \alpha^{(n-1)i_{2}}\\
\vdots & \vdots & \vdots & \vdots\\
1 & \alpha^{i_{u}} & \alpha^{2i_{u}} & \dots & \alpha^{(n-1)i_{u}}
\end{pmatrix}.\]
It is clear that $\M(A)$ is the generator matrix of $\bar{\C}_A^{\perp}$. Let $h_0$ denote a non-zero codeword in $\bar{\C}_A^{\perp}$ with the minimum weight, and
$c_{1}, c_{2}, \dots, c_{u}\in \bF_{q^d}$ be the coefficients of each row of $\M(A)$ respectively in the linear combination of
expression of~$h_{0}$. If we set $f(x) = c_{1}x^{i_{1}}+c_{2}x^{i_{2}}+\dots+c_{u}x^{i_{u}}$, then
\[h_{0}=\left( f(1), \,\, f(\alpha), \,\, f(\alpha^{2}), \,\, \dots,\,\, f(\alpha^{n-1}) \right) ,\]
and the weight of $h_{0}$ is $d_{A}^{\perp}$.

Let $B=\{ \alpha^{j_{1}}, \alpha^{j_{2}}, \dots, \alpha^{j_{v}} \}\subset \R_n$ and $\bar{\C}_B$ denote the cyclic code of length~$n$ over $\bF_{q^d}$ with the complete defining set $B$. Then $\bar{\C}_B$ has the following parity-check matrix:
\[\M(B)=
\begin{pmatrix}
1&\alpha^{j_{1}}&\alpha^{2j_{1}}&\dots&\alpha^{(n-1)j_{1}}\\
1&\alpha^{j_{2}}&\alpha^{2j_{2}}&\dots&\alpha^{(n-1)j_{2}}\\
\vdots&\vdots&\vdots&\vdots\\
1&\alpha^{j_{v}}&\alpha^{2j_{v}}&\dots&\alpha^{(n-1)j_{v}}
\end{pmatrix}.
\]
For any $e\in \{1, 2, \dots, v\}$, consider the following
matrix associated with $A \alpha^{j_e}$:
\[ \M(A \alpha^{j_{e}} )=
\begin{pmatrix}
1&\alpha^{i_{1}+j_{e}}&\alpha^{2(i_{1}+j_{e})}&\dots&\alpha^{(n-1)(i_{1}+j_{e})}\\
1&\alpha^{i_{2}+j_{e}}&\alpha^{2(i_{2}+j_{e})}&\dots&\alpha^{(n-1)(i_{2}+j_{e})}\\
\vdots&\vdots&\vdots&\vdots\\
1&\alpha^{i_{u}+j_{e}}&\alpha^{2(i_{u}+j_{e})}&\dots&\alpha^{(n-1)(i_{u}+j_{e})}
\end{pmatrix}.
\]
Using the same coefficients as in the expression of $h_{0}$ to make a linear combination of rows of $\M(A \alpha^{j_{e}})$ we get
\[h_{e}= \left( f(1), \,\, \alpha^{j_{e}} f(\alpha), \,\, \alpha^{2j_{e}}f(\alpha^{2}), \,\, \dots, \,\, \alpha^{(n-1)j_{e}}f(\alpha^{n-1}) \right), \,\, 1\leq e\leq v.
\]
Since the powers of $\alpha$ are non-zero, $h_{e}$ shares the same support with $h_{0}$, whose index set is denoted by $I=\{ s_1, s_2, \dots, s_t\}$, where $t=d_{A}^{\perp}$.

Let $H$ be a matrix whose $e$-th row is $h_{e}$ for $e=1, 2,\dots, v$, which comes from the row space of $\M(AB)$,
where $\M(AB)$ is an $|AB|\times n$ matrix from the set $AB$ as that $\M(A)$ is from $A$. It is clear that each row
of $H$ is a codeword of the dual of $\bar{\C}$. All non-zero columns of $H$ form an $v\times t$ matrix as following:
 \[H_I =\begin{pmatrix}
\alpha^{s_1j_{1}} f(\alpha^{s_1}) & \alpha^{s_2 j_{1}}f(\alpha^{s_2}) & \dots & \alpha^{s_tj_{1}}f(\alpha^{s_t}) \\
\alpha^{s_1j_{2}} f(\alpha^{s_1}) & \alpha^{s_2 j_{2}}f(\alpha^{s_2}) & \dots & \alpha^{s_tj_{2}}f(\alpha^{s_t})\\
\vdots                            &     \vdots                        & \vdots&\vdots\\
\alpha^{s_1j_{v}} f(\alpha^{s_1}) & \alpha^{s_2j_{v}} f(\alpha^{s_2}) & \dots &\alpha^{s_tj_{v}}f(\alpha^{s_t})
\end{pmatrix}.
\]
Since $\M(B)$ is a parity-check matrix of $\bar{\C}_B$, any $d_{B}-1$ columns of $\M(B)$ are linearly independent over $\bF_{q^d}$. So,
any $d_{B}-1$ columns of $H_I$ are linearly independent over $\bF_{q^d}$. By Lemma~\ref{lem:equdef} the cyclic code $\bar{\C}$ has
 $(d_{A}^{\perp}- d_B +1, d_B)$-locality.
\EOP

\begin{remark}
Let $\C_{A}$ denote the cyclic code of length~$n$ over $\bF_q$ with the complete defining set $A\subset \R_n$. Set $B=\{ 1 \}$ in Theorem~\ref{thm:locality}. Then $AB=A$ and $d_{B}=2$. From Theorem~\ref{thm:locality} we know that $\C_A$ has $(d_A^\perp-1, 2)$-locality, where $d_A^\perp$ is the minimum distance of the dual of the cyclic code of length~$n$ over $\bF_{q^d}$ with the complete defining set $A$. Since the set $\{ 0\leq j\leq n-1\, |\, \alpha^j \in A \}$ is a union of some $q$-cyclotomic cosets modulo $n$, $d_A^\perp$ is also the minimum distance of the dual of $\C_A$ by Lemma~\ref{lem:distance} and Proposition~\ref{prop:twod}. So, Theorem~\ref{thm:locality} is a generalization of $r$-locality of cyclic codes.
\end{remark}

\begin{remark}
Follow the notation in Lemma~\ref{lem:bin2} and let $B=\{1,\alpha^b,\alpha^{2b},\dots, \alpha^{(\delta-2)b}\}$ and $A=\alpha^{l_1} \left \langle \alpha^{r+\delta-1} \right \rangle$ be two subsets of $\R_n$. It is easy to verify that the set $\cup_{l}L_l$ in Lemma~\ref{lem:bin2} can be rewritten as
\[ \cup_{l=l_1}^{l_{\delta-1}}L_l = AB .\]
Let $\C_A$ denote the cyclic code of length $n$ over $\bF_q$ with the complete defining set $A$. Proposition~\ref{prop:exactd} implies that the minimum distance of the dual of $\C_A$, $d_A^{\perp} = r+\delta-1$. From the BCH bound and Singleton bound we see $d_B=\delta$. By Theorem~\ref{thm:locality}, the cyclic codes whose complete defining set
contain the set $\cup_{l=l_1}^{l_{\delta-1}}L_l$ have $(r, \delta)$-locality. So, Lemma~\ref{lem:bin2} is a special case of Theorem~\ref{thm:locality}.
\end{remark}

The following corollary provide a way to construct cyclic $(r, \delta)$-LRCs over $\bF_q$ of length $n$ via two known cyclic codes over $\bF_q$ of length $n$.

\begin{corollary}\label{cor:rdelta}
Let $\C_A$ and $\C_B$ denote two cyclic codes of length $n$ over $\bF_q$ with the complete defining sets $A, B\subset \R_n$, respectively, which satisfy $d_A^\perp > d_B$,
where $d_A^\perp$ and $d_B$ are the minimum distance of the dual of $\C_A$ and $\C_B$, respectively. Then the cyclic code $\C_{AB}$ over $\bF_q$ with the complete defining set $AB$ has $(d_A^\perp-d_B+1, d_B)$-locality.
\end{corollary}
\pf Assume that the order of $q$ modulo $n$ is $d$, then $\R_n \subset \bF_{q^d}$. Set $ K_1 =\{ 0\leq \ell\leq n-1 \,\, |\,\, \alpha^{\ell} \in A \}$ and $ K_2 = \{ 0\leq s\leq n-1\,\, |\,\, \alpha^s\in B \}$. Then $AB= \{ \alpha^i \,\, |\,\, i \equiv \ell+s \,\, {\rm mod} \,\, n, \,\, \ell \in K_1, \,\, s\in K_2 \}$. It is easy to verify that the set $\{ 0\leq i\leq n-1\,\, |\,\, \alpha^i \in AB\}$ is a union of some $q$-cyclotomic cosets modulo~$n$. So, the generator polynomial of $\C_{AB}$ is over $\bF_q$.
By Theorem~\ref{thm:locality}, we know $\C_{AB}$ has $(\bar{d}_A^\perp-\bar{d}_B+1, \bar{d}_B)$-locality, where $\bar{d}_A^\perp$ and $\bar{d}_B$ are the minimum distances of the dual of the cyclic code over $\bF_{q^d}$ with the complete defining set $A$ and the cyclic code over $\bF_{q^d}$ with the complete defining set $B$, respectively.
By Lemma~\ref{lem:distance} and Proposition~\ref{prop:twod}, $\bar{d}_A^\perp$ and $\bar{d}_B$ are also the minimum distance of the dual of the cyclic code over $\bF_{q}$ with the complete defining set $A$ and the cyclic code over $\bF_{q}$ with the complete defining set $B$, respectively.   \EOP

Below we give an example to illustrate Corollary 3.5.

\begin{example}
Let $n=31$, $q=2$ and $\alpha$ be a root of $x^5 + x^2 + 1$, which is a $31$-th primitive root of unity in $\bF_{2^{32}}$.
Let $\C_A=\langle x^6 + x^5 + x^3 + x^2 + x + 1 \rangle $ and $\C_B = \langle x^5 + x^4 + x^2 + x + 1 \rangle$ be two cyclic codes of length $31$ over $\bF_2$.
The cyclic codes $\C_A$ and $\C_B$ have the complete defining sets $A=\{\alpha^0, \alpha^1, \alpha^2, \alpha^4, \alpha^8, \alpha^{16}\}$ and
$B=\{ \alpha^5, \alpha^{10}, \alpha^{20}, \alpha^{9}, \alpha^{18}\}$ respectively. Magma shows that $d^{\perp}_A=15$, which is the minimum distance of the dual of $\C_A$,
and $d_B=3$, which is the minimum distance of $\C_B$. It is easy to see that
\[AB =\left\{ \alpha^3, \alpha^6, \alpha^{12}, \alpha^{24}, \alpha^{17}, \alpha^5, \alpha^{10}, \alpha^{20}, \alpha^9, \alpha^{18}, \alpha^7, \alpha^{14}, \alpha^{28},
\alpha^{25}, \alpha^{19}, \alpha^{11}, \alpha^{22}, \alpha^{13}, \alpha^{26}, \alpha^{21}\right\} \]
and $g(x)=\prod_{\beta\in AB}(x-\beta)=x^{20} + x^{19} + x^{17} + x^{15} + x^{14} + x^{13} + x^{10} + x^7 + x^6 + x^5 + x^3 + x + 1$.
Corollary~\ref{cor:rdelta} implies that $\langle g(x)\rangle$ is a cyclic code over $\bF_2$ and has $(13,3)$-locality.
\end{example}

\section{Optimal cyclic $(r, \delta)$-LRCs with length~$n$ for $n\,|\, (q-1)$ }\label{sec:q-1}

In this section we assume that $n\,|\, (q-1)$. In this case we know that all $n$-th roots of unity are in $\bF_q$, i.e., $\R_n \subset \bF_q$. Let $A, B$ and $AB$ be subsets of $\R_n$. In the following we denote $\C_A$, $\C_B$ and $\C_{AB}$ the cyclic codes of length~$n$ over $\bF_q$ with the complete defining sets $A$, $B$ and $AB$,
respectively, and denote $d_A^\perp$ and $d_B$ the minimum distance of the dual of $\C_A$ and $\C_B$, respectively. The $(r, \delta)$-locality of $\C_{AB}$ is known from Theorem~\ref{thm:locality}. Below we construct optimal cyclic $(r, \delta)$-LRCs of the form $\C_{AB}$ by choosing proper subsets $A, B\subset \R_n$. Specifically, we set $B=\{1,\alpha^b,\alpha^{2b},\dots,\alpha^{(\delta-2)b}\}$ for some positive integer $b$ with $(b,n)=1$. By application of the BCH bound and Betti-Sala bound~\cite{Betti-Sala2006}, we obtain optimal cyclic LRCs via choosing a proper subset $A\subset\R_n$.
In particular, we show that there exist optimal cyclic $(r, \delta)$-LRCs satisfying $(r+\delta-1) \nmid n $.

\subsection{The construction from the BCH bound }

In this subsection, we give a generic construction of optimal cyclic $(r, \delta)$-LRCs by using the BCH bound on cyclic codes.

\begin{theorem}\label{thm:optimality}
Let $\delta, b, s,n$ be positive integers and $\alpha$ be a primitive $n$-th root of unity, where $n\, |\, (q-1)$, $(b, n)=1$ and $\delta \geq 2$. Consider the following subsets of $\R_n$:
$$A= \left\{ \alpha^{t}, \alpha^{t+b},\dots,\alpha^{t+(m-1)b},\alpha^{t+i_1 b}, \alpha^{t+i_2 b},\dots,\alpha^{t+i_s b} \right\},\,
B=\left\{ 1,\alpha^b, \alpha^{2b},\dots,\alpha^{(\delta-2)b}\right\}, $$
where $t\in \{ 0,1, \dots, n-1\}$, $m-1+\delta \leq i_1< i_2<\dots< i_s \leq n-\delta$ and $i_{\ell+1}-i_\ell\geq \delta$ for $\ell=1, 2, \dots, s-1$.
Then the code $\C_{AB}$ is a cyclic $(d_{A}^{\perp}- \delta +1, \delta)$-LRC with dimension $k=n-m+1-(s+1)(\delta-1)$, where $d_{A}^{\perp}$ is the minimum distance of the dual of $\C_A$, which is the cyclic code of length $n$ over $\bF_q$ with the complete defining set $A$. Moreover, if $\lceil \frac{k}{r} \rceil=s+1$, where $r=d_{A}^{\perp}- \delta +1$, then $\C_{AB}$ is an optimal $(r, \delta)$-LRC and has the minimum distance $m+\delta-1$.
\end{theorem}
\pf Let $\C_B$ and $d_B$ denote the cyclic code with the complete defining set $B$ and the minimum distance of $\C_B$, respectively. By the BCH and Singleton bound theorems, $d_B = \delta$. Since $i_{\ell+1}-i_\ell\geq \delta$, $\R_n \setminus A$ contains a consecutive set of length at least $\delta-1$. By Lemma~\ref{lem:distance} we know $d^{\perp}_A \geq \delta$. From Theorem~\ref{thm:locality} we see that $\C_{AB}$ has $( d^{\perp}_A- \delta+1, \delta)$-locality. It is easy to show that
\[ AB=\left\{ \alpha^t, \alpha^{t+b}, \alpha^{t+2b}, \ldots, \alpha^{t+(m+\delta-3)b}\right\} \bigcup_{j=1}^s \left\{  \alpha^{t+ i_jb}, \alpha^{t + (i_j+1)b}, \ldots,\alpha^{t + (i_j+\delta-2)b}\right\}.\]
So, the dimension of $\C_{AB}$ is equal to $k=n-|AB|=n-m+1-(s+1)(\delta-1)$. When $\lceil \frac{k}{r} \rceil=s+1$, by the Singleton-like bound of LRCs, we have the minimum distance of $\C_{AB}$, $d_{AB} \leq n-k-( \lceil \frac{k}{r}\rceil-1)(\delta-1)+1 = m+\delta-1$. On the other hand, there exists a consecutive set of length $m+\delta-2$ in the complete defining set of $\C_{AB}$. From the BCH bound we get $ d_{AB}\geq m+\delta-1$.
So, $d_{AB}= m+\delta-1$ and $\C_{AB}$ is optimal.
\EOP

To construct optimal cyclic $(r, \delta)$-LRCs of length~$n$ over $\bF_q$, it is crucial to design proper subsets $A, B\subset \R_n$ such that $\lceil \frac{k}{r} \rceil=s+1$ by Theorem~\ref{thm:optimality}, where $k$ is the dimension of $\C_{AB}$ over $\bF_q$ and $r=d_A^\perp-\delta+1$. We will provide a choice of subsets $A, B\subset \R_n$ in the following corollary.

\begin{corollary}\label{cor:existenceofd}
Follow the notation in Theorem~\ref{thm:optimality} and assume $(r+\delta-1)\mid n$.
Let $\nu=n/(r+\delta-1)$ and $B=\{ \alpha^0, \alpha^b, \alpha^{2b}, \cdots, \alpha^{(\delta-2)b}\}\subset \R_n$.
Set
\begin{equation}\label{eq:cora}
 A=\left\{\begin{split}
&\alpha^{t},\alpha^{t+b}, \alpha^{t+2b}, \dots,\alpha^{t + (\ell(r+\delta-1)+i)b}, \\
&\alpha^{t+((\ell+1)(r+\delta-1)+j)b}, \alpha^{t+((\ell+2)(r+\delta-1)+j)b}, \dots,\alpha^{t+((\nu-1)(r+\delta-1)+j)b}
\end{split} \right\}, \end{equation}
where $0\leq t\leq n-1$, $0\leq i\leq r-1$ and $\ell, j$ satisfy one of the following conditions:
\begin{itemize}
\item $0\leq \ell \leq \nu-3$ and $0\leq j\leq i$;
\item $ \ell=\nu-2$  and $ j=i$.
\end{itemize}
Then $\C_{AB}$ is an optimal cyclic $(r,\delta)$-LRC of length $n$ over $\bF_q$ with dimension $(\nu-\ell)r-i$ and  minimum distance
$\delta+i+\ell(r+\delta-1)$.
\end{corollary}
\pf Take $m=\ell(r+\delta-1)+i+1$, $s=\nu-\ell-1$ and $i_t=(\ell+t)(r+\delta-1)+j$ for $1\leq t\leq s$ in $A$ of Theorem~\ref{thm:optimality}, then we know that $\C_{AB}$ has $(d_A^\perp-\delta+1, \delta)$-locality and the dimension $k=(\nu-\ell)r-i$. Next, we show that $d_A^\perp -\delta+1= r$. It is easy to see that $A$ contains a coset of a subgroup of $\R_n$ with order~$\nu$, i.e, $\alpha^{t+jb}	\left \langle \alpha^{(r+\delta-1)b} \right \rangle$ and $\R_n\setminus A$ contains a consecutive set of length $r+\delta-2$, i.e,
$$ \left\{ \alpha^{t+((\nu -2)(r+\delta-1)+j+1)b}, \alpha^{t+((\nu -2)(r+\delta-1)+j+2)b},\dots,\alpha^{t+((\nu-1)(r+\delta-1)+j-1)b} \right\}.$$
Proposition~\ref{prop:exactd} implies that $d_A^{\perp} = r+\delta-1$. Since $i<r$ we have
$$\left\lceil\frac{k}{r}\right\rceil=\left\lceil\frac{(\nu-\ell) r-i}{r}\right\rceil=\nu-\ell=s+1.$$
By Theorem~\ref{thm:optimality}, the code $\C_{AB}$ is an optimal cyclic $(r, \delta)$-LRC with minimum distance $\delta+i+\ell(r+\delta-1)$.
\EOP

\begin{remark}
Follow the notation in Lemma~\ref{lem:construction-chen} and Corollary~\ref{cor:existenceofd} and assume $r|k$. Let $\mu=\frac{k}{r}$ and $\nu=\frac{n}{r+\delta-1}$. Suppose $i=j=0$ and $\ell=\nu-\mu$ in (\ref{eq:cora}). Then $AB$ in Corollary~\ref{cor:existenceofd} is reduced to
\begin{equation}\label{eq:corab}
\begin{split}
AB=\bigcup_{e=1}^{\mu-1} \alpha^{t+(\nu-\mu+e)(r+\delta-1)b}B
\bigcup\left\{  \alpha^t, \alpha^{t+b}, \alpha^{t+2b}, \dots, \alpha^{ t+((\nu-\mu)(r+\delta-1)+\delta-2)b}\right\}.
\end{split}
\end{equation}
If $t\in L_{l_1}$, where $L_{l_1}$ is defined in Lemma~\ref{lem:construction-chen}, then $AB$ in (\ref{eq:corab}) is exact the $\left(\cup_{l}L_l\right)\cup D$ in Lemma~\ref{lem:construction-chen} by substituting $n=\nu(r+\delta-1)$ and $k=\mu r$. So, Corollary~\ref{cor:existenceofd} includes the construction in Lemma~\ref{lem:construction-chen}.
\end{remark}

When $s=1$ in Theorem~\ref{thm:optimality}, $\C_{AB}$ is an optimal cyclic $(r,\delta)$-LRC of length $n$ over $\bF_q$ provided $\lceil \frac{k}{r}\rceil=2$,
where $k$ is the dimension of $\C_{AB}$ over $\bF_q$ and $r=d_A^\perp-\delta+1$. In this case, it provides us a way to find optimal cyclic $(r, \delta)$-LRCs such that
$(r+\delta-1) \nmid n$ in the following corollary.

\begin{corollary}~\label{cor:amds}
Let $\delta, b, n$ be positive integers and $\alpha$ be a primitive $n$-th root of unity, where $n\, |\, (q-1)$, $(b, n)=1$ and $\delta \geq 2$.
Consider the following subsets of $\R_n$:
$$ A=\{\alpha^{t},\alpha^{t+b},\dots,\alpha^{t+(m-1)b},\alpha^{t+\ell b}\}, \,\, B=\{1,\alpha^b,\alpha^{2b},\dots,\alpha^{(\delta-2)b}\}, $$
where $t\in \{0,1,\dots,n-1\}$ and $m-1+\delta \leq \ell \leq n-\delta$. Then the code $\C_{AB}$ is a cyclic  $(d_{A}^{\perp}-\delta +1, \delta)$-LRC with dimension $n-m-2\delta+3$. If $\delta-2 < n-m-d_A^{\perp}$, then $\C_{AB}$ is optimal and has the minimum distance $m+\delta-1$.
\end{corollary}
\pf Take $s=1$ and $i_1=\ell$ in $A$ of Theorem~\ref{thm:optimality}, then we know that the code $\C_{AB}$ has $(d_A^\perp -\delta+1,\delta)$-locality and the dimension $k=n-m-2\delta+3$.
 To show that $\C_{AB}$ is optimal, by Theorem~\ref{thm:optimality} we need to verify that $\lceil \frac{k}{r}\rceil =2$, where $r= d_A^\perp -\delta+1$.
It is easy to verify that $d_A^{\perp} \geq \delta$ by Lemma~\ref{lem:distance} and
\[ AB = \left\{ \alpha^t, \alpha^{t+b}, \dots, \alpha^{t+(m+\delta-3)b}, \alpha^{t+\ell b},  \alpha^{t+ (\ell+1) b}, \dots,  \alpha^{t+(\ell+\delta-2) b}  \right\} .\]
By application of the BCH bound and Singleton-like bound for $\C_{AB}$, we have
\begin{equation}\label{eq:inqtyda}
m+\delta-1 \leq d_{AB} \leq n-k-\left( \left\lceil\frac{k}{r}\right\rceil -1\right)(\delta-1) +1.
\end{equation}
Since $n-k= m+2\delta-3$, from (\ref{eq:inqtyda}) we get
\[ m+\delta-1 \leq m+2\delta-3 - \left( \left\lceil \frac{k}{r}\right\rceil -1\right)(\delta-1) +1 .\]
This implies that $\lceil \frac{k}{r}\rceil \leq 2$. Thus, to show that $\lceil \frac{k}{r}\rceil =2$, it is enough to show that $k>r$.
Since $\delta-2 < n-m-d_A^{\perp}$ we have
\[ k-r = n-(m+2\delta-3) -(d_A^\perp -\delta+1)= (n-m-d_A^\perp) -(\delta-2) >0 .\]
This completes the proof.
\EOP

Below we give an example to illustrate Corollary~\ref{cor:amds}

\begin{example}
Let $n=18$ and $\alpha$ be an $18$-th primitive root of unity in $\bF_{19}$. Let
$A=\{ \alpha,\alpha^{2},\alpha^{3},\alpha^{4},\alpha^{5},\alpha^{9} \}$, and this is the case of $t=1, b=1, m=5$ and $\ell=8$ in Corollary~\ref{cor:amds}. Magma verifies that $\C_A^\perp$ has the parameters $[18,6,10]$, where $\C_A^\perp$ is the dual of the cyclic code with the complete defining set $A$. Let $d_A^\perp$ denote the minimum distance
of $\C_A^{\perp}$. It is clear that $\delta=2$, $3$ or $4$ satisfies
\[ 0 \leq \delta-2 < n-m-d_A^\perp .\]
Assume $\delta=2$ and $B=\{ 1\}$. Then
\[ AB = \left\{ \alpha, \alpha^{2}, \alpha^{3},\alpha^{4}, \alpha^{5},  \alpha^{9} \right\} ,\] and $r=d_A^\perp - \delta+1 = 9$. So,
$\C_{AB}$ is an optimal cyclic $(9, 2)$-LRC with parameters $[18, 12, 6]$.
Assume $\delta=3$ and $B=\{ 1, \alpha \}$. Then
\[ AB = \left\{ \alpha, \alpha^{2}, \alpha^{3}, \alpha^{4}, \alpha^{5}, \alpha^{6}, \alpha^{9}, \alpha^{10} \right\} ,\]
and $r=d_A^\perp - \delta+1 = 8$. So, $\C_{AB}$ is an optimal cyclic $(8, 3)$-LRC with parameters $[18, 10, 7]$.
Assume $\delta=4$ and $B=\{ 1, \alpha, \alpha^2 \}$. Then
\[ AB = \left\{ \alpha,\alpha^{2},\alpha^{3},\alpha^{4},\alpha^{5},\alpha^{6},\alpha^{7},\alpha^{9},\alpha^{10},\alpha^{11} \right\} ,\]
and $r=d_A^\perp - \delta +1 = 7$. So, $\C_{AB}$ is an optimal cyclic $(7, 4)$-LRC with parameters $[18, 8, 8]$.
In the above three cases, $r+\delta-1=10$, which doesn't divide 18.
\end{example}

\begin{remark}
Corollary~\ref{cor:amds} gives us a method to find optimal cyclic $(r,\delta)$-LRCs satisfying $(r+\delta-1)\, \nmid\, n$ for the case of $s=1$ in
Theorem~\ref{thm:optimality}, where $r=d_A^\perp-\delta+1$. When $s\geq 2$, we don't know if there exist $A, B\subset \R_n$ such that $\lceil \frac{k}{r} \rceil=s+1$ provided that $(r+\delta-1)\, \nmid\, n$, where $k$ is the dimension of $\C_{AB}$. Note that the required precondition of $\delta-2 < n-m-d_A^{\perp}$ in Corollary~\ref{cor:amds}
is not necessary. Assume that $\delta=2$, i.e., $B=\{ 1 \}$ in Corollary~\ref{cor:amds}, then $\C_A$ is always an optimal cyclic $(d_A^{\perp}-1)$-LRC although the required
precondition may not hold.
\end{remark}

\begin{corollary}\label{cor:delta=2}
Follow the notation in Corollary~\ref{cor:amds} and assume that $\delta=2$. Then $\C_{A}$ is an optimal cyclic $(d_A^\perp-1)$-LRC for any~$\ell$ with $m+1\leq \ell\leq n-2$,
where $d_A^\perp$ is the minimum distance of the dual of $\C_A$.
\end{corollary}
\pf From the BCH bound and Singleton bound, we have $m+2\geq d_A \geq m+1$, where $d_A$ is the minimum distance of $\C_A$, which is the cyclic code of length $n$ over $\bF_q$ with the complete defining set~$A$. If $d_A=m+2$, then~$\C_{A}$ is an $[n,n-(m+1),m+2]$ MDS code, and so $\C_{A}^{\perp}$ is also an $[n,m+1,n-m]$ MDS code, where $\C_{A}^{\perp}$ is the dual of $\C_A$. In this case, $r=d_A^{\perp}-1=n-m-1$ and $\lceil \frac{k}{r}\rceil=1$, where $d_A^\perp$ is the minimum distance of $\C_A^\perp$. So, $d_A$ achieves the Singleton-like bound $m+2$ and $\C_{A}$ is optimal. If $d_A=m+1$, then~$\C_{A}$ has the parameters $[n,n-(m+1),m+1]$ and $\C_{A}^{\perp}$ has the parameters $[n,m+1, d]$, where $d \leq n-m-1$. In this case, it is easy to chek that $\lceil \frac{k}{r}\rceil \geq 2$. On the other hand, from the proof of Corollary~\ref{cor:amds}, we have $\lceil \frac{k}{r}\rceil \leq 2$. So, $\lceil \frac{k}{r}\rceil=2$ and $d_A$ achieves the Singleton-like bound $m+1$. Hence, $\C_{A}$ is optimal.
\EOP

\begin{remark}
Corollary~\ref{cor:delta=2} shows that all known cyclic MDS and AMDS codes are optimal cyclic $r$-LRCs.
\end{remark}

Below we give an example to illustrate Corollary~\ref{cor:delta=2}.
\begin{example}
Let $n=31$ and $\alpha$ be a $31$-th primitive root of unity in $\bF_{32}$.  Let $\C_A$ denote the cyclic code of length $31$ over $\bF_{32}$ with the complete defining set
$A \subset \R_{31}$.
\begin{description}
\item[{\rm (1)}] Assume that $A=\{ \alpha^0,\alpha^{1},\alpha^{8}\}$. Magma figures out that $\C_A$ and its dual have the parameters $[31,28,4]$ and
$[31,3,29]$, respectively. So, $\C_A$ is a MDS code and also an optimal cyclic $28$-LRC.
\item[{\rm (2)}] Assume that $A=\{ \alpha^0,\alpha^{1},\alpha^{7}\}$. Magma figures out that $\C_A$ and its dual have the parameters $[31,28,3]$ and
$[31,3,28]$, respectively. So, $\C_A$ is an AMDS code and also an optimal cyclic $27$-LRC.
\end{description}
\end{example}

\subsection{The construction from the Betti-Sala bound}

In this subsection we provide a construction of optimal cyclic $(r, \delta)$-LRCs by application of the Betti-Sala bound on the minimum distance of cyclic codes.

\begin{theorem}\label{thm:optimality02}
Let $b, m, \delta$ be positive integers with $\delta \geq 2$ and $\alpha\in \bF_q$ be a primitive $n$-th root of unity, where $n\,|\, (q-1)$ and $(b, n)=1$.
Consider two subsets of $\R_n$:
$B=\left\{ 1, \alpha^b, \alpha^{2b}, \dots, \alpha^{(\delta-2)b} \right\}$ and
$$ A=\left\{ \alpha^{t}, \alpha^{t+b}, \dots,\alpha^{t+((m-1)\delta+1)b}, \alpha^{t+(m\delta+1)b}, \alpha^{t+((m+1)\delta+1)b}, \dots, \alpha^{t+(2m\delta+1)b} \right\}, $$
for some $t\in\{0,1,\dots,n-1\}$. Then the code $\C_{AB}$ is a cyclic $(d_{A}^{\perp}-\delta +1, \delta)$-LRC with dimension $k=n-m\delta-(m+1)(\delta-1)$, where $d_A^\perp$ is the minimum distance of the dual of the cyclic code of length $n$ over $\bF_{q}$ with the complete defining set $A$. Moreover, if $\lceil \frac{k}{r} \rceil=m+1$, where $r=d_{A}^{\perp}-\delta +1$, then $\C_{AB}$ is an optimal $(r, \delta)$-LRC and has the minimum distance $(m+1)\delta$.
\end{theorem}
\pf Let $\C_B$ and $d_B$ denote the cyclic code of length $n$ over $\bF_q$ with the complete defining set $B$ and the minimum distance of $\C_B$, respectively.
It is clear that $d_B=\delta$. We can verify that $\R_n \setminus A$ contains a consecutive set of length at least $\delta-1$. By Lemma~\ref{lem:distance} we know that $d^{\perp}_A \geq \delta$, where $d_A^\perp$ is the minimum distance of the dual of the cyclic code over $\bF_q$ with the complete defining set $A$. From Theorem~\ref{thm:locality} we know that $\C_{AB}$ has $(r, \delta)$-locality, where $r= d^{\perp}_A- \delta+1$. It is easy to see that
\[ AB =\left\{ \alpha^t, \alpha^{t+b}, \dots, \alpha^{t+(m\delta-1)b}\right\} \bigcup_{j=0}^m \left\{ \alpha^{t+ ((m+j)\delta+1) b}, \alpha^{t + ((m+j)\delta+2)b}, \dots,\alpha^{t + ((m+j)\delta + \delta-1)b}\right\}.\]
So, the dimension of $\C_{AB}$ is $k=n- m\delta-(m+1)(\delta-1)$, and $d_{AB} \geq (m+1)\delta$ from Lemma~\ref{lem:genBSbound}, where $d_{AB}$ is the minimum distance of $\C_{AB}$. When $\lceil \frac{k}{r} \rceil=m+1$, by the Singleton-like bound of LRCs, we have $d_{AB} \leq n-k-( \lceil \frac{k}{r}\rceil-1)(\delta-1)+1 = (m+1)\delta$.
So, $d_{AB}= (m+1)\delta$ and $\C_{AB}$ is optimal.
\EOP

If $m=1$ in Theorem~\ref{thm:optimality02}, then we get the following proposition.

\begin{prop}\label{prop:BSbond}
Let $b, \delta$ be positive integers and $\alpha\in \bF_q$ be a primitive $n$-th root of unity, where $n\,|\, (q-1)$ and $(b, n)=1$.
Consider two subsets of $\R_n$ as follows:
\[  A=\left\{ \alpha^{t}, \alpha^{t+b}, \alpha^{t+(\delta+1)b}, \alpha^{t+(2\delta+1)b} \right\}, \,\, B=\left\{ 1, \alpha^b, \alpha^{2b}, \dots, \alpha^{(\delta-2)b}\right\},  \]
where $0\leq t\leq n-1$. Then the code $\C_{AB}$ is a cyclic $(d_{A}^{\perp}-\delta +1, \delta)$-LRC with dimension $n-3\delta+2$. Moreover, if $d_A^\perp < n-2\delta+1$ then $\C_{AB}$ is optimal and has the minimum distance $2\delta$.
\end{prop}
\pf Take $m=1$ in $A$ of Theorem~\ref{thm:optimality02}, then we know that the code $\C_{AB}$ has $(d_A^\perp -\delta+1,\delta)$-locality and the dimension $k=n-3\delta+2$.
To show that $\C_{AB}$ is optimal, by Theorem~\ref{thm:optimality02}, we need to verify that $\lceil \frac{k}{r}\rceil =2$, where $r= d_A^\perp -\delta+1$.
It is easy to show that the set of zeros of $\C_{AB}$ is as follows:
\[ \left\{ \alpha^t, \alpha^{t+b}, \dots, \alpha^{t +(\delta-1)b}\right\}\bigcup_{i=0}^1 \left\{ \alpha^{t+((i+1)\delta+1)b}, \alpha^{t+((i+2)\delta+1)b}, \dots, \alpha^{t+((i+1)\delta+\delta-1)b}\right\} .\]
By application of the Betti-Sala bound and Singleton-like bound for $\C_{AB}$, we have
\begin{equation}\label{eq:inqtyda2}
2\delta \leq d_{AB} \leq n-k-\left( \left\lceil \frac{k}{r}\right\rceil -1\right)(\delta-1) +1,
\end{equation}
where $d_{AB}$ is the minimum distance of $\C_{AB}$. From (\ref{eq:inqtyda2}) we get
 \[ 2\delta \leq 3\delta -2-  \left( \left\lceil \frac{k}{r}\right\rceil -1\right)(\delta-1) +1 .\]
This implies that $\lceil \frac{k}{r}\rceil \leq 2$. Thus, to show that $\lceil \frac{k}{r}\rceil =2$, it is enough to show that $k>r$.
In fact, from the given condition we have
\[ k-r = n-(3\delta-2) -(d_A^\perp -\delta+1)= (n-2\delta +1 ) -d_A^\perp >0 .\]
This completes the proof.
\EOP

Below we give an example to illustrate Proposition~\ref{prop:BSbond}.

\begin{example}
Let $n=18$ and $\alpha$ be a $18$-th primitive root of unity in $\bF_{19}$. Set
$A= \left\{ 1, \alpha, \alpha^{5}, \alpha^{9} \right\}$ and  $B=\left\{ 1, \alpha, \alpha^2 \right\}$.
The dual $\C_A^{\perp}$ of the cyclic code $\C_A$ with the complete defining set $A$ has the parameters $[18,4,9]$. It is verified that
$9=d_A^\perp < n-2\delta+1$ for $n=18$ and $\delta=4$. So, the cyclic code $\C_{AB}$ with the complete defining set
$$ AB=\{1, \alpha, \alpha^2, \alpha^3, \alpha^5, \alpha^6, \alpha^7, \alpha^9, \alpha^{10}, \alpha^{11}\}$$
is an optimal cyclic $(6,4)$-LRC with parameters $[18,8,8]$.
\end{example}

If we further restrict the length of the cyclic codes discussed above, then we obtain a class of optimal cyclic LRCs
with special parameters as follows.

\begin{prop}\label{prop:n4delt+2}
Let $\delta$ be an integer with $\delta\geq 2$ and $n=4\delta+2$. Let $q$ be a prime power with $n\mid (q-1)$ and $\alpha$ be a primitive $n$-th root of unity in $\bF_q$. Set
\[ A=\left\{ \alpha^{t}, \alpha^{t+b}, \alpha^{t+(\delta+1)b}, \alpha^{t+(2\delta+1)b}\right\},\,\, B=\left\{1,\alpha^b,\alpha^{2b},\dots,\alpha^{(\delta-2)b}\right\},\]
where $t \in \{0, 1, \dots, n-1\}$ and $(b,n)=1$, then $\C_{AB}$ is an optimal cyclic $(\delta+2,\delta)$-LRC with parameters $[4\delta+2,\delta+4,2\delta]$.
\end{prop}
\pf Let $\C_A$ and $d_A^\perp$ denote the cyclic code with the complete defining set $A$ and the minimum distance of the dual of $\C_A$.
From Proposition~\ref{prop:exactd} we have $d_A^{\perp}=2\delta+1$, i.e., $\C_A^\perp$ has the parameters $[n, 4, 2\delta+1]$.
It is easy to verify that $d_A^\perp < n-2\delta+1$. By Proposition~\ref{prop:BSbond} we have that $\C_{AB}$ is an optimal $(\delta+2,\delta)$-LRC
with  parameters $[4\delta+2,\delta+4,2\delta]$.
\EOP

\section{\hspace{-4mm} Optimal cyclic $(r,\delta)$-LRCs with length~$n$ for $n\mid(q+1)$}\label{sec:q+1}

In this section we assume that $n\,|\, (q+1)$. In this case the order of $q$ modulo~$n$ is $2$, and so all $n$-th roots of unity are in $\bF_{q^2}$, i.e,
$\R_n \subset \bF_{q^2}$. For an integer $s$ with $1\leq s\leq n-1$, $sq \equiv -s \,\, {\rm mod}\,\, n $. So, the $q$-cyclotomic coset of $s$ modulo $n$ is
of the form $\{s, -s\}$, and for a cyclic code of length $n$ over $\bF_q$, $\alpha^s$ is its zero if and only if $\alpha^{-s}$ is also its zero.
Let $A$ and $B$ be subsets of $\R_n$. If $AB$ is a complete defining set of a cyclic code over $\bF_q$, then the set $\{ 0\leq j \leq n-1\,\, |\,\, \alpha^j \in AB \}$
is a union of some $q$-cyclotomic cosets modulo~$n$. In the following let $\C_{AB}$ be the cyclic code of length $n$ over $\bF_{q}$ with the complete defining set $AB\subset \bF_{q^2}$, and $d_A^\perp$ and $d_B$ denote the minimum distance of the dual of the cyclic code with the complete defining set $A$ and the cyclic code with the complete defining set $B$ over $\bF_{q^2}$, respectively unless we specify them separately. From Theorem~\ref{thm:locality} we know that $\C_{AB}$ has $(d_A^\perp-d_B+1, d_B)$-locality. Below we construct optimal cyclic $(r, \delta)$-LRCs of the form $\C_{AB}$ over $\bF_q$ with lengths $q+1$ and its factors by choosing proper subsets $A, B\subset \R_n$. Our constructions include all optimal cyclic $(r, \delta)$-LRCs proposed in~\cite{Chen2018,Chen2019}. Moreover, many optimal cyclic $(r, \delta)$-LRCs of length $n\, |\, (q+1)$ such that
$(r+\delta-1) \nmid n$ can be obtained from our method.

\subsection{Optimal cyclic $(r, \delta)$-LRCs with even $\delta$}\label{subsec:evendelta}

In this subsection we first give a generic criterion to construct optimal cyclic $(r, \delta)$-LRCs of length $n\, |\, (q+1)$
for the case of $\delta$ being even. Then several explicit constructions of optimal cyclic $(r, \delta)$-LRCs are proposed.

\begin{theorem}\label{thm:evendelta}
Let $\delta, b, s, m$ be positive integers and $\alpha$ be a primitive $n$-th root of unity, where $n\mid (q+1)$, $(b, n)=1$ and $\delta$ is even. Consider the following subsets of $\R_n$:
\[A= \left\{ \alpha^{t}, \alpha^{t+b},\dots,\alpha^{t+(m-1)b},\alpha^{t+i_1 b}, \alpha^{t+i_2 b},\dots,\alpha^{t+i_s b} \right\}, \,\,
B=\left\{ \alpha^0,\alpha^{\pm b},\dots,\alpha^{\frac{\pm (\delta-2)b}{2}}\right\}, \]
where $t\in \{ 0,1, \dots, n-1\}$, $m-1+\delta \leq i_1< i_2<\dots< i_s \leq n-\delta $ and $i_{\ell+1}-i_\ell\geq \delta$ for $\ell=1, 2, \dots, s-1$.
If the set $\{ 0\leq j\leq n-1 \,|\, \alpha^j \in AB\}$ is a union of some $q$-cyclotomic cosets modulo~$n$, then $\C_{AB}$ is a cyclic $(d_{A}^{\perp}- \delta +1, \delta)$-LRC of length $n$ over $\bF_q$ with dimension $k=n-m+1-(s+1)(\delta-1)$, where $d_A^\perp$ denote the minimum distance of the dual of the cyclic code over $\bF_{q^2}$ with the complete defining set $A$. Moreover, if $\lceil \frac{k}{r} \rceil=s+1$, where $r= d_{A}^{\perp}- \delta +1$, then $\C_{AB}$ is an optimal $(r,\delta)$-LRC and has the minimum distance $m+\delta-1$.
\end{theorem}
\pf Let $g(x) = \prod_{\beta\in AB} (x-\beta)$. Since the set $\{ 0\leq j \leq n-1 \, |\, \alpha^j \in AB \}$ is a union of some $q$-cyclotomic cosets modulo $n$, $g(x)$ is a polynomial over $\bF_q$. Let $\C_{AB}$ and $\bar{\C}_{AB}$ denote the cyclic codes of length~$n$ generated by $g(x)$ over $\bF_{q}$ and $\bF_{q^2}$, respectively. It is clear that the dimension of $\C_{AB}$ over $\bF_q$ is equal to that of $\bar{\C}_{AB}$ over $\bF_{q^2}$. Theorem~\ref{thm:optimality} implies that $\bar{\C}_{AB}$ is an optimal cyclic $(d_A^\perp-\delta+1,\delta)$-LRC of length $n$ over $\bF_{q^2}$ with dimension $n-m+1-(s+1)(\delta-1)$ and minimum distance $m+\delta-1$. So, the dimension of $\C_{AB}$ over $\bF_q$ is also $n-m+1-(s+1)(\delta-1)$. Since $\C_{AB}$ is the subfield subcode of $\bar{\C}_{AB}$, $\C_{AB}$ also has the minimum distance $d_{AB}=m+\delta-1$ by Proposition~\ref{prop:twod} and the same $(d_A^\perp-\delta+1,\delta)$-locality by Lemma~\ref{lem:localityofsubfiledsubcode}. So, $d_{AB}$ achieves the Single-like bound and
$\C_{AB}$ is an optimal cyclic $(d_A^\perp-\delta+1,\delta)$-LRC with dimension $n-m+1-(s+1)(\delta-1)$ and minimum distance $m+\delta-1$.
\EOP

To construct optimal cyclic $(r, \delta)$-LRCs of the form $\C_{AB}$ over $\bF_q$ with lengths $q+1$ and its factors, it is crucial to find the proper subsets $A, B\subset \R_n$ satisfying some conditions of Theorem~\ref{thm:evendelta}. The following corollary gives three classes of explicit representations of $A$ and $B$
to construct optimal cyclic $(r, \delta)$-LRCs with flexible minimum distance. Many optimal cyclic $(r, \delta)$-LRCs can be obtained
by the choice of some parameters.

\begin{corollary}\label{cor:ndivq+1X}
Let $n$, $r$ and $\delta$ be positive integers with $(n,q)=1$, $n\mid (q+1)$, $(r+\delta-1)\mid n$ and $\delta$ even.
Let $\nu= \frac{n}{r+\delta-1}$ and $B=\{\alpha^0,\alpha^{\pm 1},\dots,\alpha^{\frac{\pm(\delta-2)}{2}}\}\subset \R_n$ .
\begin{description}
\item[(1)] Let
\begin{equation}\label{eq:case1}
 A =\left\{\begin{aligned}
 &\alpha^0, \,\, \, \alpha^{\pm 1}, \,\, \dots,\,\,  \alpha^{\pm(\ell(r+\delta-1)+i)}, \\ &\alpha^{(\ell+1)(r+\delta-1)},\alpha^{(\ell+2)(r+\delta-1)},\dots,\alpha^{(\nu-\ell-1)(r+\delta-1)}
\end{aligned}\right\},
\end{equation}
where $i \in \{ 0, 1,\dots, \lfloor\frac{r-1}{2}\rfloor\}$ and $\ell\in\{ 0, 1, \dots, \lfloor\frac{\nu-2}{2}\rfloor \}$.
Then $\C_{AB}$ is an optimal cyclic $(r,\delta)$-LRC over $\bF_{q}$ with dimension $(\nu-2\ell)r-2i$ and minimum distance $\delta+2i+2\ell(r+\delta-1)$,
\item[(2)] When $r+\delta-1$ is even, let
\begin{equation}\label{eq:case2}
 A=\left\{
\begin{aligned}
&\alpha^0, \,\, \alpha^{\pm 1},\,\, \dots, \,\,\alpha^{\pm \left(\frac{2\ell+1}{2}(r+\delta-1)+i\right)}, \\
&\alpha^{\frac{2\ell+3}{2}(r+\delta-1)}, \alpha^{\frac{2\ell+5}{2}(r+\delta-1)},\dots,\alpha^{\frac{2\nu-2\ell-3}{2}(r+\delta-1)}
\end{aligned} \right\},\end{equation}
where $i \in \{0,1,\dots, \lfloor\frac{r-1}{2}\rfloor\}$ and $\ell\in\{ 0, 1, \dots, \lfloor\frac{\nu-3}{2}\rfloor\}$.
Then $\C_{AB}$ is an optimal cyclic $(r,\delta)$-LRC over $\bF_{q}$  with dimension $(\nu-2\ell-1)r-2i$ and minimum distance~$\delta+2i+(2\ell+1)(r+\delta-1)$.
\item[(3)] When $\nu$ is odd, let
\begin{equation}\label{eq:case3}
A=\left\{\begin{aligned} \alpha^{(\frac{\nu-1}{2}-\ell)(r+\delta-1)-i}, \alpha^{(\frac{\nu-1}{2}-\ell)(r+\delta-1)-i+1}, \dots,\alpha^{(\frac{\nu+1}{2}+\ell)(r+\delta-1)+i}, \\
\alpha^{(\frac{\nu+1}{2}+\ell+1)(r+\delta-1)}, \alpha^{(\frac{\nu+1}{2}+\ell+2)(r+\delta-1)},\dots,\alpha^{(\frac{3\nu-1}{2}-\ell-1)(r+\delta-1)}
\end{aligned} \right\}, \end{equation}
where $i \in \{0,1,\dots,\lfloor\frac{r-1}{2}\rfloor\}$ and $\ell\in\{1,2,\dots, \frac{\nu-3}{2}\}$. Then $\C_{AB}$ is an optimal cyclic $(r,\delta)$-LRC over $\bF_{q}$ with dimension $(\nu-2\ell-1)r-2i$ and minimum distance~$\delta+2i+(2\ell+1)(r+\delta-1)$.
\end{description}
\end{corollary}
\pf (1) Let $\C_A$ and $d_A^\perp$ denote the cyclic code of length $n$ over $\bF_{q^2}$ with the complete defining set $A$ and the minimum distance of the dual of $\C_A$, respectively. We can verify that $A$ contains a coset of a subgroup of $\R_n$ with order~$\nu$ and $\R_n\setminus A$ contains a consecutive set of length $r+\delta-2$. So,
$d_A^{\perp} = r+\delta-1$ by Proposition~\ref{prop:exactd}. Set $t=-\ell(r+\delta-1)-i$, $b=1$, $m=1+2i+2\ell(r+\delta-1)$, $s=\nu-2\ell-1$ and $i_j=(2\ell+j)(r+\delta-1)+i$ for $j=1,2, \dots, s$ in $A$ of Theorem~\ref{thm:evendelta}, then we get the representation of $A$ in (\ref{eq:case1}). It is not hard to verify that
\begin{equation*}\label{eq:ab}
AB=\bigcup_{j=1}^{\nu-2\ell-1} \alpha^{(\ell+j)(r+\delta-1)}B\bigcup\left\{  \alpha^0, \alpha^{\pm 1}, \alpha^{\pm 2}, \dots, \alpha^{\pm \left( \ell(r+\delta-1)+i +\frac{\delta-2}{2}\right)}\right\}
\end{equation*}
is a disjoint union and the set $\{ 0\leq j\leq n-1 \, |\, \alpha^j \in AB \}$ is a union of some $q$-cyclotomic cosets modulo~$n$. So, the dimension of $\C_{AB}$ is equal to $k=n-|AB|=(\nu-2\ell) r-2i$ and $\lceil\frac{k}{r}\rceil=\lceil\frac{(\nu-2\ell) r-2i}{r}\rceil=\nu-2\ell=s+1$. Hence, Theorem~\ref{thm:evendelta} implies that $\C_{AB}$ is an optimal cyclic $(r, \delta)$-LRC with minimum distance $\delta+2i+2\ell(r+\delta-1)$.

(2) By a similar discussion in the case~(1), we have that $d_A^{\perp} = r+\delta-1$ from Proposition~\ref{prop:exactd}. Set $t=-\frac{2\ell+1}{2}(r+\delta-1)-i$, $b=1$,
$m=1+2i+(2\ell+1)(r+\delta-1)$, $s=\nu-2\ell-2$ and $i_j=(2\ell+1+j)(r+\delta-1)+i$ for $j=1,2, \dots, s$ in $A$ of Theorem~\ref{thm:evendelta}. Then we get the representation of $A$ in (\ref{eq:case2}). It is not hard to verify that
\begin{equation*}\label{eq:ab}
AB=\bigcup_{j=1}^{\nu-2\ell-2} \alpha^{\frac{2\ell+1+2j}{2}(r+\delta-1)}B\bigcup\left\{  \alpha^0, \alpha^{\pm 1}, \alpha^{\pm 2}, \dots, \alpha^{\pm \left( \frac{2\ell+1}{2}(r+\delta-1)+i+\frac{\delta-2}{2}\right)}\right\}
\end{equation*}
is a disjoint union and the set $\{ 0\leq j\leq n-1 \, |\, \alpha^j \in AB \}$ is a union of some $q$-cyclotomic cosets modulo~$n$. So, the dimension of $\C_{AB}$ is equal to $k=n-|AB|=(\nu-2\ell-1) r-2i$ and $\lceil\frac{k}{r}\rceil=\lceil\frac{(\nu-2\ell-1) r-2i}{r}\rceil=\nu-2\ell-1=s+1$. Hence, Theorem~\ref{thm:evendelta} implies that $\C_{AB}$ is an optimal cyclic $(r, \delta)$-LRC over $\bF_q$ with minimum distance $\delta+2i+(2\ell+1)(r+\delta-1)$.

(3) Similar to the discussion of the case~(1), we obtain that $d_A^{\perp} = r+\delta-1$ from Proposition~\ref{prop:exactd}. Set $t=(\frac{\nu-1}{2}-\ell)(r+\delta-1)-i$, $b=1$, $m=1+2i+(2\ell+1)(r+\delta-1)$, $s=\nu-2\ell-2$ and $i_j=(2\ell+1+j)(r+\delta-1)+i$ for $j=1,2, \dots, s$ in $A$ of Theorem~\ref{thm:evendelta}. Then we get the representation of $A$ in (\ref{eq:case3}). It is not hard to verify that
\[\begin{split}\label{eq:ab}
AB=&\bigcup_{j=1}^{\nu-2\ell-2} \alpha^{(\frac{\nu+1}{2}+\ell+j)(r+\delta-1)}B \\
&\bigcup\left\{  \alpha^{(\frac{\nu-1}{2}-\ell)(r+\delta-1)-i-\frac{\delta-1}{2}}, \alpha^{(\frac{\nu-1}{2}-\ell)(r+\delta-1)-i-\frac{\delta-1}{2}+1}, \dots,\alpha^{(\frac{\nu+1}{2}+\ell)(r+\delta-1)+i+\frac{\delta-1}{2}}\right\}
\end{split}\]
is a disjoint union and the set $\{ 0\leq j\leq n-1 \, |\, \alpha^j \in AB \}$ is a union of some $q$-cyclotomic cosets modulo~$n$. So, the dimensionof $\C_{AB}$ is equal to $k=n-|AB|=(\nu-2\ell-1) r-2i$ and $\lceil\frac{k}{r}\rceil=\lceil\frac{(\nu-2\ell-1) r-2i}{r}\rceil=\nu-2\ell-1=s+1$. Hence, Theorem~\ref{thm:evendelta} implies that $\C_{AB}$ is an optimal cyclic $(r, \delta)$-LRC over $\bF_q$ with minimum distance $\delta+2i+(2\ell+1)(r+\delta-1)$. \EOP

\begin{remark}
It is not hard to verify that the set $\{ 0\leq j\leq n-1\,\, |\,\, \alpha^j \in A \}$ in the above three cases is a union of some $q$-cyclotomic cosets modulo~$n$. By Lemma~\ref{lem:distance} and Proposition~\ref{prop:twod}, $d_A^\perp$ in the above cases is also the minimum distance of the dual of the cyclic code of length $n$ over $\bF_q$ with the complete defining set~$A$. It is known that the common differences of arithmetic progression in the set $\{ 0\leq j\leq n-1\,\, |\,\, \alpha^j \in A \}$ in Corollary~\ref{cor:ndivq+1X} are $1$ and $r+\delta-1$, which can be replaced by $b$ and $b(r+\delta-1)$, respectively for some positive integer $b$ with $(b, n)=1$.
\end{remark}

\begin{remark}
In the following we investigate the relation between the constructions in Corollary~\ref{cor:ndivq+1X} and those in~\cite{Chen2018,Chen2019}.
It is known that $\C_{AB}$ in the case (1) is an optimal cyclic $(r, \delta)$-LRC  and has the parameters $[n,\, (\nu-2\ell)r-2i, \, \delta+2i+2\ell(r+\delta-1)]$, where $\nu=\frac{n}{r+\delta-1}$, $i \in \{ 0, 1,\dots, \lfloor\frac{r-1}{2}\rfloor\}$ and $\ell\in\{ 0, 1, \dots, \lfloor\frac{\nu-2}{2}\rfloor \}$.
Assume that $r\, |\, k$ and let $\mu=\frac{k}{r}$, $i=0$ and $\ell=\frac{\nu-\mu}{2}$. Then $AB$ in the case~(1) is reduced to
\begin{equation}\label{eq:ab}
AB=\bigcup_{j=1}^{\mu-1} \alpha^{(\frac{\nu-\mu}{2}+j)(r+\delta-1)}B\bigcup\left\{  \alpha^0, \alpha^{\pm 1}, \alpha^{\pm 2}, \dots, \alpha^{\pm \left( \frac{\nu-\mu}{2}(r+\delta-1)+\frac{\delta-2}{2}\right)}\right\}.
\end{equation}
It is verified that $AB$ in (\ref{eq:ab}) is exact the $\left(\cup_{l}L_l\right)\cup D$ in Constructions~37, 45 and 49 in~\cite{Chen2018} by substituting $n=\nu(r+\delta-1)$ and $k=\mu r$. So, the case (1) of Corollary~\ref{cor:ndivq+1X} includes those constructions in~\cite{Chen2018}.
Similarly, the case (2) of Corollary~\ref{cor:ndivq+1X} includes the construction in Theorem~1 of~\cite{Chen2019} and the
case (3) includes Constructions~35, 40 and~47 in~\cite{Chen2018}. Our method seems more convenient to obtain optimal cyclic $(r, \delta)$-LRCs with flexible parameters
by the choice of different values of~$i$ in $A$ of Corollary~\ref{cor:ndivq+1X}.
\end{remark}

Below we give some examples to illustrate Corollary~\ref{cor:ndivq+1X}.

\begin{example}
Let $q=23$, $n=24=q+1$, $r=3$ and $\delta=4$. Then $r+\delta-1=6$ and $\nu=\frac{n}{r+\delta-1}=4$. So, $r+\delta-1$ and
$\nu$ are both even. Let $B=\{\alpha^{-1}, \alpha^0, \alpha^1\}$.
\begin{description}
\item[(1)] Let $i=\ell=1$ in $A$ of the case~(1). Then $A=\left\{ \alpha^0, \alpha^{\pm 1}, \dots, \alpha^{\pm 7}, \alpha^{12}\right\},$
So, $\C_{AB}$ is an optimal cyclic $(3,4)$-LRC over $\bF_{23}$ with parameters $[24,4,18]$.
\item[(2)] Let $i=1$ and $\ell=0$ in $A$ of the case (2). Then
$A=\left\{ \alpha^{0}, \alpha^{\pm 1}, \dots, \alpha^{\pm 4}, \alpha^{9}, \alpha^{15} \right\}.$
So, $\C_{AB}$ is an optimal cyclic $(3,4)$-LRC over $\bF_{23}$ with parameters $[24,7,12]$.
\end{description}
\end{example}

\begin{example}
Let $q=7^2=49$, $n=50=q+1$, $r=5$ and $\delta=6$. Then $r+\delta-1=10$, $\nu=\frac{n}{r+\delta-1}=5$. So,
$r+\delta-1$ is even and $\nu$ is odd. Let $B=\{ \alpha^0, \alpha^{\pm 1}, \alpha^{\pm 2}\}$.
\begin{description}
\item[(1)] Let $i=\ell=1$ in $A$ of the case (1). Then $A=\{ \alpha^{0}, \alpha^{\pm 1}, \dots, \alpha^{\pm 11}, \alpha^{20},
\alpha^{30}\}$. So, $\C_{AB}$ is an optimal cyclic $(5,6)$-LRC over $\bF_{49}$ with parameters $[50,13,28]$.
\item[(2)] Let $i=\ell=1$ in $A$ of the case (2). Then $A=\{ \alpha^{0}, \alpha^{\pm 1}, \dots, \alpha^{\pm 16}, \alpha^{25}\}$.
So, $\C_{AB}$ is an optimal cyclic $(5,6)$-LRC over $\bF_{49}$ with parameters $[50,8,38]$.
\item[(3)] Let $i=\ell=1$ in $A$ of the case (3). Then $A=\{ \alpha^{9}, \alpha^{10}, \alpha^{11},\dots, \alpha^{41}, \alpha^{0}\}$.
So, $\C_{AB}$ is an optimal cyclic $(5,6)$-LRC over $\bF_{49}$ with parameters $[50,8,38]$.
\end{description}
\end{example}

\begin{example}
Let $q=2^6$, $n=65=q+1$, $r=2$ and $\delta=4$. Then $r+\delta-1=5$ and $\nu=\frac{n}{r+\delta-1}=13$. So, $r+\delta-1$ and $\nu$
are both odd. Let $B=\{ \alpha^0, \alpha^{\pm 1}\}$.
\begin{description}
\item[(1)] Let $i=1$ and $\ell=3$ in $A$ of the case (1). Then $A=\{\alpha^{0}, \alpha^{\pm 1}, \dots, \alpha^{\pm 16}, \alpha^{20},\alpha^{25},\\
\alpha^{30},\alpha^{35},\alpha^{40},\alpha^{45}\}$. So, $\C_{AB}$ is an optimal cyclic $(5,6)$-LRC over $\bF_{64}$ with parameters $[65,12,36]$.
\item[(2)] Let $i=1$ and $\ell=3$ in $A$ of the case (3). Then $ A=\{\alpha^{14},\alpha^{15}, \alpha^{16}, \dots,\alpha^{51},
\alpha^{55},\alpha^{60},\\ \alpha^{0},\alpha^{5},\alpha^{10}\}$. So, $\C_{AB}$ is an optimal cyclic $(5,6)$-LRC over $\bF_{64}$ with parameters $[65,10,41]$.
\end{description}
\end{example}

In Theorem~\ref{thm:evendelta}, let $s=1$ in $A$, then it is possible to construct optimal cyclic $(r, \delta)$-LRCs of length $n\, |\, (q+1)$ such that $(r+\delta-1) \nmid n$ as the following corollary.

\begin{corollary}~\label{cor:evendelta}
Let $b, m$ and $n$ be positive integers such that $n$ is odd, $m$ is even, $n\, |\, (q+1)$ and $(b, n)=1$.
Let $\delta$ be an even integer with $ 2\leq \delta \leq \frac{n-m+1}{2}$. Consider the following subsets of $\R_n$:
$$ A=\left\{\alpha^{0},\alpha^{\pm \frac{n+1}{2}b},\alpha^{\pm \frac{n+3}{2}b},\dots,\alpha^{\pm \frac{n+m-1}{2}b}\right\}, \, \,\, B=\left\{\alpha^0,\alpha^{\pm b},\dots,\alpha^{\frac{\pm (\delta-2)b}{2}}\right\}.$$
Then $\C_{AB}$ is a cyclic $(d_{A}^{\perp}- \delta +1, \delta)$-LRC with dimension $n-m-2\delta+3$, where $d_A^\perp$ is the minimum distance of the dual of the cyclic code of length $n$ over $\bF_{q^2}$ with the complete defining set $A$. If $\delta-2 < n-m-d_A^{\perp}$, then $\C_{AB}$ is optimal and has the minimum distance $m+\delta-1$.
\end{corollary}
\pf Take $s=1$, $t=\frac{n-m+1}{2}b$, $i_1=\frac{n+m-1}{2}$ and note that $\alpha^{-\frac{n+i}{2}}=\alpha^{\frac{n-i}{2}}$ for $1\leq i\leq m-1$ in $A$ of Theorem~\ref{thm:evendelta}, then we get the set $A$ in Corollary~\ref{cor:evendelta}. It is easy to verify that
$$ AB=\left\{\alpha^0,\alpha^{\pm b},\dots,\alpha^{\frac{\pm (\delta-2)b}{2}},\alpha^{\pm \frac{n+1}{2}b},\alpha^{\pm \frac{n+3}{2}b},\dots,\alpha^{\pm \frac{n+m+\delta-3}{2}b}\right\},$$
and the set $\{ 0\leq j\leq n-1 \, |\, \alpha^j \in AB \}$ is a union of some $q$-cyclotomic cosets modulo~$n$.
By Theorem~\ref{thm:evendelta}, $\C_{AB}$ has $(d_{A}^{\perp}-\delta +1, \delta)$-locality and the dimension $k=n-m-2\delta+3$.
Since $\delta-2 < n-m-d_A^{\perp}$, by a similar discussion in Corollary~\ref{cor:amds} we have $\lceil \frac{k}{r} \rceil=2$, where $r=d_{A}^{\perp}- \delta +1$.
By Theorem~\ref{thm:evendelta}, $\C_{AB}$ is an optimal cyclic $(r, \delta)$-LRC with minimum distance $m+\delta-1$.
\EOP

Below we give an example to illustrate Corollary~\ref{cor:evendelta}.

\begin{example}
Let $n=33$, $m=6$ and $\alpha$ be an $33$-th primitive root of unity in $\bF_{32^2}$. Let
$A=\{ 1,\alpha^{14},\alpha^{15} ,\alpha^{16} ,\alpha^{17} ,\alpha^{18},\alpha^{19} \}$. Magma verifies the dual code $\C_A^\perp$ of the cyclic code with
the complete defining set $A$ over $\bF_{32^2}$ has the parameters $[33,7,23]$. It is clear that $\delta=2$ or $4$  satisfies
$0 \leq \delta-2 < n-m-d_A^\perp$, where $d_A^\perp = 23$.

Assume $\delta=2$ and set $B=\{ 1 \}$. Then
\[ AB = \left\{1,\alpha^{14},\alpha^{15} ,\alpha^{16} ,\alpha^{17} ,\alpha^{18},\alpha^{19} \right\} \]
and $r=d_A^\perp - \delta+1 = 22$. So, $\C_{AB}$ is an optimal cyclic $(22, 2)$-LRC over $\bF_{32}$ with  parameters $[33, 26, 7]$.

Assume $\delta=4$ and set $B=\{ \alpha^{-1},1,\alpha^1 \}$. Then
\[ AB = \left\{\alpha^{-1},1,\alpha^1,\alpha^{13},\alpha^{14},\alpha^{15} ,\alpha^{16} ,\alpha^{17} ,\alpha^{18},\alpha^{19},\alpha^{20}  \right\} \]
and $r=d_A^\perp - \delta+1 = 20$. So, $\C_{AB}$ is an optimal cyclic $(20, 4)$-LRC over $\bF_{32}$ with  parameters $[33, 22, 9]$.

In the above two cases, $r+\delta-1=23$, which doesn't divide $33$.
\end{example}

\begin{remark}
In Theorem~\ref{thm:evendelta}, when $s>1$ in $A$, we don't know if there exist $A, B\subset \R_n$ such that $\lceil \frac{k}{r} \rceil=s+1$ provided that $(r+\delta-1)\, \nmid\, n$, where $k$ is the dimension of $\C_{AB}$. In the case of $n$ being even, we have a similar result as Corollary~\ref{cor:evendelta}, but we can't find a numeric example of optimal cyclic $(r,\delta)$-LRCs ($\delta\geq 2$) such that $(r+\delta-1)\nmid n$. So, we omit the statement. Furthermore, if $\delta=2$ in Corollary~\ref{cor:evendelta}, then all the cyclic code with the complete defining set $A$ is an optimal cyclic $(d_A^\perp-1)$-LRC by
the similar proof of Corollary~\ref{cor:delta=2}.
\end{remark}

\subsection{Optimal cyclic $(r, \delta)$-LRCs with odd $\delta$}\label{subsec:odddelta}

By a similar discussion in Subsection~\ref{subsec:evendelta} we consider the construction of optimal cyclic
$(r, \delta)$-LRCs for odd $\delta$. First, we give a generic criterion to construct optimal cyclic $(r, \delta)$-LRCs,
then an explicit construction is proposed for the case of $\delta$ being odd. The proofs in this subsection are similar to
those in Subsection~\ref{subsec:evendelta}. So, we omit the details here.
\begin{theorem}\label{thm:odddelta}
Let $\delta, b, s, m$ be positive integers and $\alpha$ be a primitive $n$-th root of unity, where $n\mid (q+1)$, $(b, n)=1$ and $\delta$ is odd. Consider the following subsets of $\R_n$:
\[
\begin{split}
A &= \left\{ \alpha^{t}, \alpha^{t+b},\dots,\alpha^{t+(m-1)b},\alpha^{t+i_1 b}, \alpha^{t+i_2 b},\dots,\alpha^{t+i_s b} \right\}, \\
B &=\left\{ \alpha^{-\frac{\delta-3}{2}b},\dots,\alpha^{-b},\alpha^0,\alpha^b,\dots,\alpha^{\frac{\delta-1}{2}b}\right\},
\end{split} \]
where $t\in \{ 0,1, \dots, n-1\}$, $m-1+\delta \leq i_1< i_2<\dots< i_s \leq n-\delta $ and $i_{\ell+1}-i_\ell\geq \delta$ for $\ell=1, 2, \dots, s-1$.
If the set $\{ 0\leq j\leq n-1 \, |\, \alpha^j \in AB \}$ is a union of some $q$-cyclotomic cosets modulo~$n$, then $\C_{AB}$ is a cyclic $(d_{A}^{\perp}- \delta +1, \delta)$-LRC of length $n$ over $\bF_q$ with dimension $k=n-m+1-(s+1)(\delta-1)$, where $d_A^\perp$ denote the minimum distance of the dual of the cyclic code over $\bF_{q^2}$ with the complete defining set $A$. Moreover, if $\lceil \frac{k}{r} \rceil=s+1$, where $r= d_{A}^{\perp}- \delta +1$, then $\C_{AB}$ is an optimal $(r, \delta)$-LRC and has the minimum distance $m+\delta-1$.
\end{theorem}

\begin{corollary}\label{cor:ndivq+1Y}
Let $n$, $r$ and $\delta$ be positive integers with $(n,q)=1$, $n\mid (q+1)$, $(r+\delta-1)\mid~n$ and $\delta$ odd.
Let $B=\{\alpha^{-\frac{\delta-3}{2}},\dots,\alpha^1,\alpha^0,\alpha^1,\dots,\alpha^{\frac{\delta-1}{2}}\}\subset \R_n$ and $\nu= \frac{n}{r+\delta-1}$.
\begin{description}
\item[(1)] Let $r+\delta-1$ be odd and
\begin{equation}\label{eq:oddcase1}
\hspace{-6mm} A=\left\{\begin{aligned} &\alpha^{-\frac{r+\delta}{2}-\ell(r+\delta-1)-i}, \alpha^{-\frac{r+\delta}{2}-\ell(r+\delta-1)-i+1}, \dots,\alpha^{\frac{r+\delta-2}{2}+\ell(r+\delta-1)+i}, \\
&\alpha^{\frac{r+\delta-2}{2}+(\ell+1)(r+\delta-1)}, \alpha^{\frac{r+\delta-2}{2}+(\ell+2)(r+\delta-1)}, \dots,\alpha^{\frac{r+\delta-2}{2}+(\nu-\ell-2)(r+\delta-1)}
\end{aligned}\right\},
\end{equation}
where $i \in \{0,1,\dots, \lfloor\frac{r-1}{2}\rfloor\}$ and $\ell\in\{ 0, 1, \dots, \lfloor\frac{\nu-3}{2}\rfloor\}$.
Then $\C_{AB}$ is an optimal cyclic $(r,\delta)$-LRC over $\bF_{q}$  with dimension $(\nu-2\ell-1)r-2i$ and minimum distance~$\delta+2i+(2\ell+1)(r+\delta-1)$.
\item[(2)] Let $r+\delta-1$ and $\nu$ be odd, and
\begin{equation}\label{eq:oddcase2}
\hspace{-4mm} A =\left\{ \begin{aligned}
 &\alpha^{\frac{n-1}{2}-\ell(r+\delta-1)-i}, \alpha^{\frac{n-1}{2}-\ell(r+\delta-1)-i+1}, \dots, \alpha^{\frac{n-1}{2}+\ell(r+\delta-1)+i},\\ &\alpha^{\frac{n-1}{2}+(\ell+1)(r+\delta-1)},\alpha^{\frac{n-1}{2}+(\ell+2)(r+\delta-1)},\dots,\alpha^{\frac{n-1}{2}+(\nu-\ell-1)(r+\delta-1) }
\end{aligned}\right\},
\end{equation}
where $i \in \{0,1,\dots,\lfloor\frac{r-1}{2}\rfloor\}$ and $\ell\in\{1,2,\dots, \frac{\nu-3}{2}\}$. Then $\C_{AB}$ is an optimal cyclic $(r,\delta)$-LRC over $\bF_{q}$ with dimension $(\nu-2\ell)r-2i$ and minimum distance~$\delta+2i+2\ell(r+\delta-1)$.
\end{description}
\end{corollary}

\begin{remark}
Note that the common differences of arithmetic progression in the set $\{ 0\leq j \leq n-1\,|\,\alpha^j \in A\}$ in Corollary~\ref{cor:ndivq+1Y} are $1$ and $r+\delta-1$, which can be replaced by $b$ and $b(r+\delta-1)$, respectively for some positive integer $b$ with $(b, n)=1$. In the following we investigate the relation between the constructions in Corollary~\ref{cor:ndivq+1Y} and those in~\cite{Chen2018,Chen2019}.

If $n$ in the case~(1) is odd, then $(2, n)=1$ and we replace $\alpha$ with $\alpha^2$. Then the sets $A$ and $B$ in the case~(1) can be rewritten as
$ B=\{\alpha^{-(\delta-3)},\dots,\alpha^{-2},\alpha^0,\alpha^2,\dots,\alpha^{(\delta-1)}\}$ and
\[A
=\left\{\begin{aligned} &\alpha^{-(r+\delta)-2\ell(r+\delta-1)-2i}, \alpha^{-(r+\delta)-2\ell(r+\delta-1)-2i+2}, \dots,\alpha^{(r+\delta-2)+2\ell(r+\delta-1)+2i}, \\
&\alpha^{(r+\delta-2)+2(\ell+1)(r+\delta-1)}, \alpha^{(r+\delta-2)+2(\ell+2)(r+\delta-1)}, \dots,\alpha^{(r+\delta-2)+2(\nu-\ell-2)(r+\delta-1)}
\end{aligned}\right\}.\]
Assume $r|k$. Let $\mu=\frac{k}{r}$, $i=0$ and $\ell=\frac{\nu-\mu-1}{2}$. Then $AB$ in the case~(1) is reduced to
\begin{equation}\label{eq:abnodd}
\begin{split}
AB=&\bigcup_{j=1}^{\mu-1} \alpha^{(r+\delta-2)+(\nu-\mu-1+2j)(r+\delta-1)}B \\
&\bigcup\left\{  \alpha^0, \alpha^{\pm 1}, \alpha^{\pm 2}, \dots, \alpha^{\pm \left( (\nu-\mu)(r+\delta-1)+\delta-2\right)}\right\}.
\end{split}
\end{equation}
It is verified that $AB$ in (\ref{eq:abnodd}) is exact the $\left(\cup_{l}L_l\right)\cup D$ in Construction~54 in~\cite{Chen2018}
by substituting $n=\nu(r+\delta-1)$ and $k=\mu r$.

When $n$ in the case~(1) is even, let $\mu=\frac{k}{r}$, $i=0$ and $\ell=\frac{\nu-\mu-1}{2}$. It is verify that $AB$ in the case~(1) is reduced to \begin{equation}\label{eq:abneven}
\begin{split}
AB=&\bigcup_{j=1}^{\mu-1} \alpha^{\frac{r+\delta-2+(\nu-\mu-1+2j)(r+\delta-1)}{2}}B \\
&\bigcup\left\{  \alpha^0, \alpha^{\pm 1}, \alpha^{\pm 2}, \dots, \alpha^{\pm \left( \frac{(\nu-\mu)(r+\delta-1)+\delta-2}{2}\right)}\right\}.
\end{split}
\end{equation}
One can show that $AB$ in (\ref{eq:abneven}) is exact the $\alpha^{\frac{n}{2}}\left(\cup_{l}L_l\right)\cup D$ by substituting $n=\nu(r+\delta-1)$ and $k=\mu r$, where $\left(\cup_{l}L_l\right)\cup D$ is the complete defining set in the construction of Theorem~2 in~\cite{Chen2019}.
So, the case (1) of Corollary~\ref{cor:ndivq+1Y} includes Construction 54 in~\cite{Chen2018} and the construction of Theorem~2 in~\cite{Chen2019}. Similarly, the case (2) of Corollary~\ref{cor:ndivq+1Y} includes Construction~53 in~\cite{Chen2018}. By using our method, it seems more convenient to obtain optimal cyclic $(r, \delta)$-LRCs with flexible parameters by the choice of different values of~$i$ in $A$ of Corollary~\ref{cor:ndivq+1Y}.
\end{remark}

Below we give some examples to illustrate Corollary~\ref{cor:ndivq+1Y}.

\begin{example}
Let $q=5^3$, $n=42 \, \mid\, (q+1)$, $r=5$ and $\delta=3$. Then $r+\delta-1=7$, $\nu=\frac{n}{r+\delta-1}=6$.
Let $B=\{\alpha^{0}, \alpha^1 \}$.
\begin{description}
 \item[(1)] Let $i=1$ and $\ell=0$ in $A$ of the case (1). Then $A=\{\alpha^{-5}, \alpha^{\pm 4}, \alpha^{\pm 3}, \alpha^{\pm 2}, \alpha^{\pm 1}, \alpha^{0}, \alpha^{10}, \\ \alpha^{17}, \alpha^{24}, \alpha^{31}\}$.
 So, $\C_{AB}$ is an optimal cyclic $(5,3)$-LRC over $\bF_{125}$ with parameters $[42,23,12]$.
 \item[(2)] Let $i=1$ and $\ell=1$ in $A$ of the case (1). Then $A=\{\alpha^{-12}, \alpha^{\pm 11}, \alpha^{\pm 10}, \alpha^{\pm 9}, \alpha^{\pm 8},\alpha^{\pm 7}, \\ \alpha^{\pm 6}, \alpha^{\pm 5}, \alpha^{\pm 4}, \alpha^{\pm 3},  \alpha^{\pm 2}, \alpha^{\pm 1}, \alpha^0, \alpha^{17}, \alpha^{24}\}$.
 So, $\C_{AB}$ is an optimal cyclic $(5,3)$-LRC over $\bF_{125}$ with parameters $[42,13,26]$.
 \end{description}
\end{example}

\begin{example}
Let $q=2^6=64$, $n=65=q+1$, $r=9$ and $\delta=5$. Then $r+\delta-1=13$, $\nu=\frac{n}{r+\delta-1}=5$.
Let $B=\{\alpha^{-1}, \alpha^0, \alpha^{1}, \alpha^{2}\}$.
\begin{description}
\item[(1)] Let $i=\ell=1$ in $A$ of the case~(1). Then $ A= \{\alpha^{-21}, \alpha^{-20}, \alpha^{-19}, \dots, \alpha^{19}, \alpha^{20}, \alpha^{32} \}$.
So, $\C_{AB}$ is an optimal cyclic $(9,5)$-LRC over $\bF_{64}$ with parameters $[65,16,46]$.
\item[(2)]Let $i=\ell=1$ in $A$ of the case (2). Then $ A= \{\alpha^{18},\alpha^{19}, \alpha^{20}, \dots, \alpha^{45}, \alpha^{46}, \alpha^{58}, \alpha^6 \}$.
So, $\C_{AB}$ is an optimal cyclic $(9,5)$-LRC over $\bF_{64}$ with parameters $[65,25,33]$.
 \end{description}
\end{example}

By a similar discussion in Corollaries~\ref{cor:amds} and~\ref{cor:evendelta} we have the following
corollary, which provides a way to find optimal cyclic $(r, \delta)$-LRCs of length $n\, |\, (q+1)$
such that $(r+\delta-1) \nmid n$ for $\delta$ being odd.

\begin{corollary}~\label{cor:odddelta}
Let $b, m$ and $n$ be positive integers such that $n$ is odd, $m$ is even, $n\, |\, (q+1)$ and $(b, n)=1$.
Let $\delta$ be an odd integer with $ 2\leq \delta \leq \frac{n-m+1}{2}$. Consider the following subsets of $\R_n$:
$$ A=\left\{\alpha^{-\frac{m}{2}b},\alpha^{-\frac{m-2}{2}b},\dots,\alpha^{\frac{m-2}{2}b},\alpha^{\frac{n-1}{2}b} \right\},\,\, B=\left\{\alpha^{-\frac{\delta-3}{2}b},\alpha^{-\frac{\delta-1}{2}b},\dots,\alpha^{-b},\alpha^0,\alpha^b,\dots,\alpha^{\frac{\delta-1}{2}b} \right\}. $$
Then the code $\C_{AB}$ is a cyclic $( d_{A}^{\perp}- \delta +1, \delta)$-LRC with dimension $n-m-2\delta+3$, where $d_{A}^{\perp}$ is the minimum distance of the dual of the cyclic code of length $n$ over $\bF_{q^2}$ with the complete defining set $A$. If $\delta-2 < n-m-d_A^{\perp}$, then $\C_{AB}$ is optimal and has the minimum distance $m+\delta-1$.
\end{corollary}

Below we give an example to illustrate Corollary~\ref{cor:odddelta}.

\begin{example}
Let $n=17$, $m=6$ and $\alpha$ be an $17$-th primitive root of unity in $\bF_{16^2}$. Let
$A=\{ \alpha^{-3},\alpha^{-2},\alpha^{-1},\alpha^{0} ,\alpha^{1} ,\alpha^{2} ,\alpha^{8} \}$. Magma verifies that
the dual $\C_A^\perp$ of the cyclic code with the complete defining set $A$ over $\bF_{16^2}$ has the parameters $[17,7,9]$.
It is clear that $\delta=3$ satisfies $0 \leq \delta-2 < n-m-d_A^\perp$, where $d_A^\perp = 9$ is the minimum distance of $\C_A^\perp$.
Assume that $\delta=3$ and $B=\{ 1,\alpha\}$. Then
\[ AB = \left\{\alpha^{-3},\alpha^{-2},\alpha^{-1},\alpha^{0} ,\alpha^{1} ,\alpha^{2} ,\alpha^{3} ,\alpha^{8} ,\alpha^{9}  \right\} ,\]
and $r=d_A^\perp - \delta+1 = 7$. So, $\C_{AB}$ is an optimal cyclic $(7, 3)$-LRC over $\bF_{16}$ with parameters $[17, 8, 8]$.
In this case, $r+\delta-1=9$, which doesn't divide $17$.
\end{example}

\section{Concluding remark }\label{sec:concluding}
In this paper, we characterized $(r,\delta)$-locality of cyclic codes from the product of two zero sets. Based on this result,
we proposed many constructions of optimal cyclic $(r,\delta)$-LRCs of length $n$ for $n\,|\,(q-1)$ or $n\,|\,(q+1)$, respectively
via the product of two sets of zeros. Our constructions include all optimal cyclic $(r,\delta)$-LRCs proposed in~\cite{Chen2018,Chen2019}.
By application of our method, many optimal cyclic $(r,\delta)$-LRCs satisfying $(r+\delta-1)\nmid n$ can be obtained.


\end{document}